\documentclass[amssymb,twocolumn,aps]{revtex4-2}
\usepackage{graphicx}
\usepackage{enumerate}
\usepackage{amssymb}
\usepackage{amsmath}
\usepackage{amsfonts}
\usepackage{color}
\usepackage{mathrsfs}

\begin{document}

\title{Quantum interaction of sub-relativistic aloof electrons with mesoscopic samples}

\author{A. Ciattoni$^1$}
\email{alessandro.ciattoni@spin.cnr.it}
\affiliation{$^1$CNR-SPIN, c/o Dip.to di Scienze Fisiche e Chimiche, Via Vetoio, 67100 Coppito (L'Aquila), Italy}

\date{\today}

\begin{abstract}
Relativistic electrons experience very slight wave packet distortion and negligible momentum recoil when interacting with nanometer-sized samples, as a consequence of the ultra-short interaction time. Accordingly, modeling fast electrons as classical point-charges provides extremely accurate theoretical predictions of energy-loss spectra. Here we investigate the aloof interaction of nanometer-sized electron beams of few keV with micron-sized samples, a regime where the classical description generally fails due to significant wavefunction broadening and momentum recoil. To cope with these effects, we use macroscopic quantum electrodynamics to analytically derive a generalized expression for the electron energy loss probability which accounts for recoil. Quantum features of the interaction are shown to get dramatically strong as the interaction length is increased and/or the electron kinetic energy is decreased. Moreover, relatively large values of the energy loss probability are found at higher energy losses and larger impact parameters, a marked quantum effect which is classically forbidden by the evanescent profile of the field produced by a moving point-charge.
\end{abstract}
\maketitle

\section{Introduction}
Inelastic scattering of fast electrons has emerged as an essential technique for characterizing the optical response of nanostructures \cite{Eger1} through the detection of their optical excitations \cite{deAb1}. The relevance of such spectroscopic technique is manly due to the combination of exceptional spatial and energy resolutions \cite{Eger2} nowadays achieved in microscopes and it also stems from the clear interpretation of electron energy-loss spectra by means of the photonic local density of state \cite{deAb2}, i.e. the sample electromagnetic Green's tensor projected along the electron propagation direction. The simplicity of such theoretical modeling hinges upon the weakness of the electron-matter coupling over the ultra-short interaction time (relativistic electron velocity and nanometric interaction length) which enables both to adopt the geometrical optics description of the unperturbed fast electron wave packet \cite{Ritc1} and to neglect momentum recoil \cite{Ritc2}. Therefore, even a narrow electron beam can be regarded as a stream of uniformly moving classical charges producing a field, described by the sample Green's tensor, which acts back on the charges by slightly decreasing their kinetic energy. Consequently, the limiting point-like description of the electron turns out to be very accurate for narrow beams and it provides the classical dielectric response theory which has proven to be a solid framework \cite{deAb3,deAb4} to interpreting the observed electron energy loss spectra produced by low-loss valence excitations, in the vast majority of the considered interaction geometries and setups \cite{Garc1,Howi1,Zaba1,Wals1,Stock,Hyun1,Kone1,Echa1}.

Predictions of the classical dielectric response theory generally worsen for low electron kinetic energies or large interaction lengths. Indeed a narrow electron beam in such regime can undergo a significant quantum distortion while interacting with the sample owing both to the transverse broadening of the unperturbed wave-packet and to the momentum exchange with the sample. Slow-electrons with sub-keV kinetic energy have been harnessed to perform holography \cite{Finkk} in the point-projection  microscopy \cite{Voge1,Mulle,Voge2}, a coherent imaging technique particularly suitable for biological samples due to the effective absence of radiation damage \cite{Latyc}. In addition, sub-relativistic electrons have shown to support generalized Kapitza–Dirac \cite{Kozak} and Smith-Purcell \cite{Tsess} effects where electron momentum recoil allows the optical grating to strongly modulate electrons in the former and the periodic structure to trigger novel photoemission processes in the latter. Large interaction lengths can be achieved by means of mesoscopic samples in the aloof configuration \cite{Tale1} where dynamic momentum interchange between relativistic electrons and the specimen has been theoretically predicted and experimentally observed \cite{Tale1}. In all these setups the electron no-recoil approximation breaks down \cite{Tale2} and, with the aim of relaxing it, suitable Maxwell–Lorentz \cite{Tale3,Tale4} and Maxwell–Schrodinger \cite{Tale5,Tale6,Tale7} numerical frameworks have been denveloped. We stress that each of these situations either deals with low electron energies or large interaction lengths so that, to the best of our knowlege, the regime where both conditions are met is still unexplored.

In this paper, we fill this gap by investigating the interaction of sub-relativistic narrow electron beams with mesoscopic samples and we discuss the ensuing quantum features of the inelastic scattering. In order to achieve micron-sized interaction lenghts we focus on the aloof configuration to prevent absorption of the electron beam by the sample. Aloof beam energy loss spectroscopy is a well known and enstablished techique \cite{Warma,Cohe1,Eche1,Itsko,Croz1,Allen}, remarkably not producing  radiation damage of the sample \cite{Kriva,Eger3,Rezzz}, and its theoretical interpretation has hitherto been  based almost exclusively on the classical dielectric response theory. An exception is provided by Ref. \cite{Cohe1} where electron recoil is not neglected but the analysis is restricted to fast (relativistic) electrons probing nanosamples \cite{Eche1}. Here we resort to macroscopic quantum electrodynamics \cite{Grune,Schee,Buhma}, recently applied to modeling electron-matter interaction \cite{Rive1,DiGi1,Hayun,DiGi2,Kfirr,Meche,Ciat1}, and we relax the standard no-recoil approximation by retaining 
the quadratic momentum part in the free electron Hamiltonian. As a consequence, our model accounts for possibly significant reshaping of the electron beam while interacting with the sample and, accordingly, it encompasses the description of low velocities and large interaction lenghts regime. We evaluate the energy loss probability of an electron scattered by an arbitrary mesoscopic sample and we compare it with its no-recoil counterpart \cite{deAb1} (to which it self-consistently reduces in the relativistic limit), their mutual discrepancy providing a quantitative assessement of the quantum trait of the interaction. We show that, in the considered regime, energy loss probability is generally much larger than the one predicted by classical response theory, and that their discrepancy gets globally stronger as the interaction length increases and/or the electron kinetic energy decreases. We also prove that the above discrepancy dramatically gets more pronounced at higher energy losses and larger impact parameters to the point that optical excitations of higher frequencies can be detected even if the electron beam axis is very distant from the sample. It is worth stressing that this is a marked quantum effect produced by the electron wavefunction reshaping during the interaction since in the classical description, in this conditions, the coupling of the electron with the sample is almost absent due to the evanescent tail of the field produced by a classical point-charge.

\section{Energy loss probability}
Consider an electron traveling in vacuum outside an arbitrarily shaped macroscopic sample whose dielectric permittivity in the frequency domain is $\varepsilon _\omega (\bf{r})$. Since the sample is here made of a generally dispersive medium, the quantized field coupled to it can be conveniently described through macroscopic quantum electrodynamics \cite{Schee} whose Hamiltonian is 
\begin{equation}
\hat H_{em} =\int {d\xi } \,\hbar \omega \,\,\hat f^\dag  \left( \xi  \right)\hat f\left( \xi  \right)
\end{equation}
where $\xi  = \left( {{\bf{r}},j,\omega } \right)$, $\int {d\xi }  = \int {d^3 {\bf{r}}} \sum\nolimits_{j = 1}^3 {\int_0^{  \infty } {d\omega } }$ and $\hat f^\dag  \left( \xi  \right)$ and $\hat f \left( \xi  \right)$ are bosonic creation and annihilation operators for the dipolar excitation (quasiparticle) at the point $\bf r$, parallel to the Cartesian unit vector ${\bf e}_j$ and of frequency $\omega$.   Choosing the Coulomb gauge, the scalar and vector potential operators in the region outside the sample (where the electron wave function lies) are $\hat \Phi \left( {\bf{R}} \right) = 0$ and 
\begin{equation} \label{Apot}
{\hat{\bf A}}\left( {\bf{R}} \right) = \int {d\xi } \left[ {\frac{\omega }{{c^2 }}\sqrt {\frac{\hbar }{{\pi \varepsilon _0 }}{\rm Im}\;\varepsilon _\omega  \left( {\bf{r}} \right)} \;{\mathcal G}_\omega  \left( {{\bf{R}},{\bf{r}}} \right){\bf{e}}_j } \right]\hat f\left( \xi  \right) + {\rm h.c.},
\end{equation}
where ${\mathcal G}_\omega  \left( {{\bf{R}},{\bf{r}}} \right)$ is the Green tensor of the sample placed in vacuum. In order to deal with highly collimated electron beams routinely used in microscopes, we here focus on paraxial electron states whose momentum distribution is strongly peaked around a main momentum ${\bf{P}}_0$, i.e. $| {{\hat{\bf P}} - {\bf{P}}_0 } | \ll P_0$ where $\hat{\bf P}$ is the electron momentum operator. As detailed in Appendix A,  the joint dynamics of the paraxial electron and the quantized field is described by the Hamiltonian $\hat H = \hat H_{em}  + \hat H_e + \hat H_{{\mathop{\rm int}} }$ where 
\begin{eqnarray} \label{HeHint}
\hat H_e  &=& V\left( {\hat P_{\rm v}  + \frac{{\hat P_{\rm t}^2 }}{{2P_0 }}} \right), \nonumber \\ 
 \hat H_{{\mathop{\rm int}} }  &=& \int {d\xi } \left[ {C(  {\hat{\bf{R}},\xi } )\hat f\left( \xi  \right) + C^* ( {\hat{\bf{R}},\xi } )\hat f^\dag  \left( \xi  \right)} \right] 
\end{eqnarray}
are the free electron and interaction Hamiltonians. Here 
${\bf{V}} = c{\bf{P}}_0 /\sqrt {m^2 c^2  + P_0^2 }$ is the main electron velocity, $\hat{\bf{P}}_{\rm t}$ and $\hat P_{\rm v} {\bf{u}}_{\rm v}$ are the parts of the operator $\hat{\bf P}$ transverse and parallel to the velocity unit vector ${\bf{u}}_{\rm v}  = {\bf{V}}/V$ (see the inset of Fig.1(a)), $C\left( {{\bf{R}},\xi } \right) = \hbar \omega \sqrt {\frac{{4\alpha }}{{c^3 }}{\rm Im} \;\varepsilon _\omega  \left( {\bf{r}} \right)} \;{\bf{V}} \cdot {\mathcal G}_\omega  \left( {{\bf{R}},{\bf{r}}} \right){\bf{e}}_j$ and $\alpha  = e^2 /\left( {4\pi \varepsilon _0 \hbar c} \right) \simeq 1/137$ is the fine-structure constant. Note that, in addition to the usual term $V\hat P_{\rm v}$ routinely used to model relativistic electron dynamics over ultra-fast interaction times \cite{Meche}, the free electron Hamiltonian $\hat H_e$ in the first of Eqs.(\ref{HeHint}) also displays the contribution $V \hat P_{\rm t}^2 / \left( {2P_0 } \right)$ which describes momentum recoil. Even though this term is much smaller than $V\hat P_{\rm v}$ in paraxial approximation, its effect is here magnified by the larger interaction times pertaining the regime we are considering in this paper.

In a typical scattering setup, the incident electron beam is nearly monoenergetic so that we choose the initial electron state $\left| {\psi _0 } \right\rangle$ in such a way that the mean value of $\hat H_e$ is $E_0  \simeq VP_0$ with uncertainty $\Delta E \ll E_0$ and that the normalization $\left\langle {{\psi _0 }} \mathrel{\left | {\vphantom {{\psi _0 } {\psi _0 }}} \right. \kern-\nulldelimiterspace} {{\psi _0 }} \right\rangle  = 1$ survives in the limit $\Delta E \to 0^ +$. As detailed in Appendix B, the wave function of such state is
\begin{equation} \label{psi0}
\psi _0 \left( {\bf{R}} \right) = \theta \left( {\frac{\ell }{2} - \left| {R_{\rm v} } \right|} \right)\frac{1}{{\sqrt \ell  }}e^{\frac{{iE_0 }}{{\hbar V}}R_{\rm v} } \phi \left( {\bf{R}} \right)
\end{equation}
where $\theta \left( \zeta  \right)$ is the Heaviside step function, $\ell  = 2\hbar V/\Delta E$ is the length of the quantization box,  ${\bf{R}}_{\rm t}$ and $R_{\rm v} {\bf{u}}_{\rm v}$ are the transverse and longitudinal parts of the position vector $\bf R$ and 
\begin{equation} \label{phi}
\phi \left( {\bf{R}} \right) = \int {d^2 {\bf{R}}'_{\rm t} } {\mathcal F}\left( {{\bf{R}}_{\rm t}  - {\bf{R}}'_{\rm t} \left| {R_{\rm v} } \right.} \right)\phi \left( {{\bf{R}}'_{\rm t} } \right),
\end{equation}
is the slowly varying envelope, where 
\begin{equation}
{\mathcal F}\left( {{\bf{R}}_{\rm t} \left| {R_{\rm v} } \right.} \right) = \frac{1}{{i\pi }}\left( {\frac{{P_0 }}{{2\hbar R_{\rm v} }}} \right)e^{i\left( {\frac{{P_0 }}{{2\hbar R_{\rm v} }}} \right)R_{\rm t}^2 }
\end{equation}
is the Fresnel kernel. Note that the evelope $\phi \left( {\bf{R}} \right)$ is determined by its profile $\phi \left( {{\bf{R}}_{\rm t} } \right)$ at the waist plane $R_{\rm v}=0$ and self-consistency is ensured by the fact that ${\mathcal F} \left( {{\bf{R}}_{\rm t} \left| {R_{\rm v} } \right.} \right) \to \delta \left( {{\bf{R}}_{\rm t} } \right)$ for $R_{\rm v} \rightarrow 0$. After labelling with $\Delta R_{\rm t} \left( {R_{\rm v} } \right)$ the transverse width of the envelope $\phi ({\bf R})$ at the plane $R_{\rm v}$, the paraxial condition turns into $\Delta R_{\rm t} \left( 0 \right) \gg \hbar /\left( {2P_0 } \right)$ and the relation $\Delta R_{\rm t} \left( {R_{\rm v} } \right) = \Delta R_{\rm t} \left( 0 \right)\sqrt {1 + \left( {R_{\rm v} /\Lambda } \right)^2 }$ shows that the beam broadens along its longitudinal direction with broadening length $\Lambda  = P_0 \left[ {\Delta R_{\rm t} \left( 0 \right)} \right]^2 /\hbar$. In the region $|R_{\rm v}| \ll \Lambda$ where beam broadening is insignificant, the envelope is practically undistorted, i.e. 
\begin{equation} \label{undist}
\phi \left( {\bf{R}} \right) \simeq \phi \left( {{\bf{R}}_{\rm t} } \right)
\end{equation}
(see Appendix B) and the wave function in Eq.(\ref{psi0}) coincides with the one generally used to model fast electrons \cite{Ritc1}. Moreover, in the region $|R_{\rm v}| > \Lambda$ and $\left| {{\bf{R}}_{\rm t} } \right| \gg \Delta R_{\rm t} \left( 0 \right)$, i.e. the lateral sides of the broadened beam, the envelope displays the unimodular factor $\exp \left\{ {\frac{i}{2}\left( {\Lambda /R_{\rm v} } \right)\left[ {R_{\rm t} /\Delta R_{\rm t} \left( 0 \right)} \right]^2 } \right\}$ producing oscillations of the wave function which get more rapid at larger distances from the beam axis (see Appendix B).

Before scattering, the electron is prepared in the state ${\left| {\psi _0 } \right\rangle }$ whereas the field is in its vacuum state. The electron-sample interaction  produces a state displaying electron-field entanglement which, in the weak coupling regime, involves only single field excitations (see the second of Eqs.(\ref{HeHint})). Accordingly, in each elementary scattering process, part of the initial electron energy $E_0$ is delivered to create a field-matter quasiparticle of energy $\hbar \omega  = E_0  - E>0$ (where $E$ is the energy of the scattered electron in the process) and, as detailed in Appendix C, the overall electron momentum resolved energy loss probability (MELP) is 
\begin{eqnarray} \label{MoEnLoPr}
 \frac{{d \mathcal P}}{{d^2 {\bf{P}}_{\rm t} d\omega}} &=& \frac{{4\alpha }}{{c }}\int {d^3 {\bf{R}}_1 } \int {d^3 {\bf{R}}_2 } \frac{{e^{-\frac{i}{\hbar }{\bf{P}}_{\rm t}  \cdot \left( {{\bf{R}}_{\rm 1t}  - {\bf{R}}_{\rm 2t} } \right)} }}{{\left( {2\pi \hbar } \right)^2 }} \cdot \nonumber  \\ 
&\cdot& e^{  i\left( {\frac{\omega }{V} + \frac{{P_{\rm t}^2 }}{{2\hbar P_0 }}} \right)\left( {R_{\rm 1v}  - R_{\rm 2v} } \right)} \phi  \left( {{\bf{R}}_1 } \right)\phi ^* \left( {{\bf{R}}_2 } \right) \cdot \nonumber \\
&\cdot& {\mathop{\rm Im}\nolimits} \left[ {{\bf{u}}_{\rm v}  \cdot {\mathcal G}_\omega  \left( {{\bf{R}}_1 ,{\bf{R}}_2 } \right){\bf{u}}_{\rm v} } \right].
\end{eqnarray}
The integration of this espression over the transverse momentum ${\bf P}_{\rm t}$ produces the Fresnel kernel $
F\left( {{\bf{R}}_{\rm 1t}  - {\bf{R}}_{\rm 2t} \left| {R_{1v}  - R_{2v} } \right.} \right)$ inside the integral so that we eventually get the energy loss probability (ELP)
\begin{eqnarray} \label{EnLoPr}
 \frac{{d{\mathcal P}}}{{d \omega}} &=& \frac{{4\alpha }}{{c }}\int {d^3 {\bf{R}}_1 } \int {d^3 {\bf{R}}_2 } e^{i\frac{\omega }{V}\left( {R_{\rm 1v}  - R_{\rm 2v} } \right)} \cdot \nonumber   \\ 
&\cdot& {\mathcal F} \left( {{\bf{R}}_{\rm 1t}  - {\bf{R}}_{\rm 2t} \left| {R_{\rm 1v}  - R_{\rm 2v} } \right.} \right)\phi \left( {{\bf{R}}_1 } \right)\phi ^* \left( {{\bf{R}}_2 } \right) \cdot \nonumber   \\ 
&\cdot& {\mathop{\rm Im}\nolimits} \left[ {{\bf{u}}_{\rm v}  \cdot {\mathcal G}_\omega  \left( {{\bf{R}}_1 ,{\bf{R}}_2 } \right){\bf{u}}_{\rm v} } \right],
\end{eqnarray}
which is the main result of this paper. 

In order to discuss the basic features of Eq.(\ref{EnLoPr}), note that the sample is fully represented by its Green tensor ${\mathcal G}_\omega  \left( {{\bf{R}}_1 ,{\bf{R}}_2 } \right)$ whose effect is not negligible only over the region  $\left| {R_{\rm 1v} } \right|,\left| {R_{\rm 2v} } \right| < L$ where $L$ is the interaction length, this enabling the restriction of the corresponding integration domains. Now, if the interaction length is much smaller than the beam broadening length, the envelopes $\phi \left( {{\bf{R}}_1 } \right)$ and $\phi \left( {{\bf{R}}_2 } \right)$ can be replaced by their waist profiles $\phi \left( {{\bf{R}}_{\rm 1t} } \right)$ and $\phi \left( {{\bf{R}}_{\rm 2t} } \right)$ in the integrand of Eq.(\ref{EnLoPr}) (see Eq.(\ref{undist})) while the Fresnel kernel ${\mathcal F} \left( {{\bf{R}}_{\rm 1t}  - {\bf{R}}_{\rm 2t} \left| {R_{\rm 1v}  - R_{\rm 2v} } \right.} \right)$ behaves as the delta function $\delta \left( {{\bf{R}}_{\rm 1t}  - {\bf{R}}_{\rm 2t} } \right)$ (see the discussion before Eq.(\ref{APPundist}) of Appendix B). Therefore we conclude that for $L \ll \Lambda$, Eq.(\ref{EnLoPr}) reduces to
\begin{eqnarray} \label{EnLoPrCl}
\frac{{d{\mathcal P}_{\rm nr} }}{{d\omega }} &=& \frac{{4\alpha }}{c}\int {d^2 {\bf{R}}_{\rm t} } \left| {\phi \left( {{\bf{R}}_{\rm t} } \right)} \right|^2 \int {dR_{\rm 1v} } \int {dR_{\rm 2v} } e^{i\frac{\omega }{V}\left( {R_{\rm 1v}  - R_{\rm 2v} } \right)} \cdot \nonumber \\
&\cdot& {\mathop{\rm Im}\nolimits} \left[ {{\bf{u}}_{\rm v}  \cdot {\mathcal G}_\omega  \left( {{\bf{R}}_{\rm t}  + R_{\rm 1v} {\bf{u}}_{\rm v} ,{\bf{R}}_{\rm t}  + R_{\rm 2v} {\bf{u}}_{\rm v} } \right){\bf{u}}_{\rm v} } \right]
\end{eqnarray}
which, for $\left| {\phi \left( {{\bf{R}}_{\rm t} } \right)} \right|^2  \approx \delta \left( {{\bf{R}}_{\rm t} } \right)$, is the well-known expression for the ELP obtained from the classical dielectric response theory \cite{deAb1} and which we hereafter label with the subscript $\rm nr$ (for no-recoil). This proves that Eq.(\ref{EnLoPr}) is more general than Eq.(\ref{EnLoPrCl})  and enables the investigation of the regime $L > \Lambda$ where quantum features arise in the ELP as a consequence of electron momentum recoil. It is worth noting that the Fresnel kernel plays a crucial role in our approach since, stemming from the quadratic term $\left( {V/2P_0 } \right)\hat P_{\rm t}^2$ of the electron Hamiltonian, it fully describes momentum recoil and, accordingly, it disappears in the relativistic limit since ${\mathcal F} \left( {{\bf{R}}_{\rm t} \left| {R_{\rm v} } \right.} \right) \to \delta \left( {{\bf{R}}_{\rm t} } \right)$ for $P_0 \rightarrow +\infty$. Such a physical interpretation of the Fresnel kernel elucidates that momontum recoil plays a double role in our approach: it produces the broadening of the unperturbed electron beam (due to ${\mathcal F} \left( {{\bf{R}}_{\rm t}  - {\bf{R}}'_{\rm t} \left| {R_{\rm v} } \right.} \right)$ in Eq.(\ref{phi})) and it accounts for the momentum exchange with the sample (due to ${\mathcal F} \left( {{\bf{R}}_{\rm 1t}  - {\bf{R}}_{\rm 2t} \left| {R_{\rm 1v}  - R_{\rm 2v} } \right.} \right)$ in Eq.(\ref{EnLoPr})).

As a final remark, we note that the ELP in Eq.(\ref{EnLoPr}) evidently increases as the interaction length $L$ increases to the point of invalidating our first-order perturbation treatment which is restricted by the condition ${\mathcal P} = \int {d\omega ({{d{\mathcal P}}}/{{d\omega }}}) \ll 1$ for the total ELP. Even though this situation goes beyond the findings of this paper, we stress that a suitable nonperturbative refinement of our treatment would provide the theoretical description of the electron-matter strong coupling triggered by very long samples and structurally affected by electron momentum recoil.

\begin{figure}
\centering
\includegraphics[width = 1\linewidth]{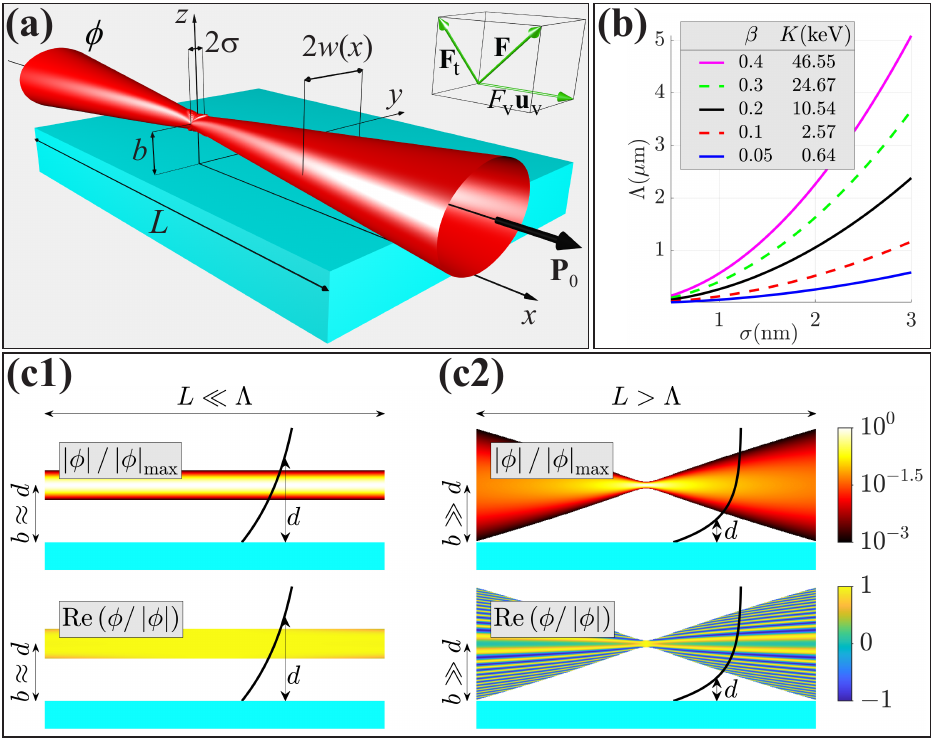}
\caption{ (a) Geometry of the interaction between an aloof broadening electron beam (red) and a mesoscopic sample (blue) (here a Gaussian beam and a planar slab, respectively, for illustrative purposes). The upper right inset displays the decomposition of a vector $\bf F$ into its parts ${{\bf F}}_{\rm t}$ transverse and $F_{\rm v}  {\bf{u}}_{\rm v}$ parallel to the beam axis. (b) Plot of the electron beam broadening length $\Lambda$ of Eq.(\ref{LAMBDA}) as function of the waist radius $\sigma$ for various electron velocities $\beta$ and corresponding kinetic energies $K$. (c) Normalized amplitude $\left| \phi  \right|/\left| \phi  \right|_{\max }$ and phase cosine ${\mathop{\rm Re}\nolimits} \left( {\phi /\left| \phi  \right|} \right)$ of the Gaussian envelope along the longitudinal plane $y=0$. Solid black curves pictorially sketch the evanescent tail of sample optical excitations of vacuum decay length $d$. (c1) For $L \ll \Lambda$ (true for nanometer-sized relativistic electron beams probing nano-samples) amplitude and phase of $\phi$ are almost undistorted within the interaction volume and the interaction admits a classical description. The optical excitation can be detected if $b \approx d$, where $b$ is the beam impact parameter. (c2) For $L > \Lambda$ (true for nanometer-sized sub-relativistic electron beams probing mesoscopic samples) quantum broadening and related side oscillations of $\phi$ forbid a classical description of the interaction. Quantum features are more distinct in the detection of the optical excitation as the ratios $L/\Lambda$ and $b/d$ increase.}
\label{Fig1}
\end{figure}

\begin{figure*}
\centering
\includegraphics[width = 1\linewidth]{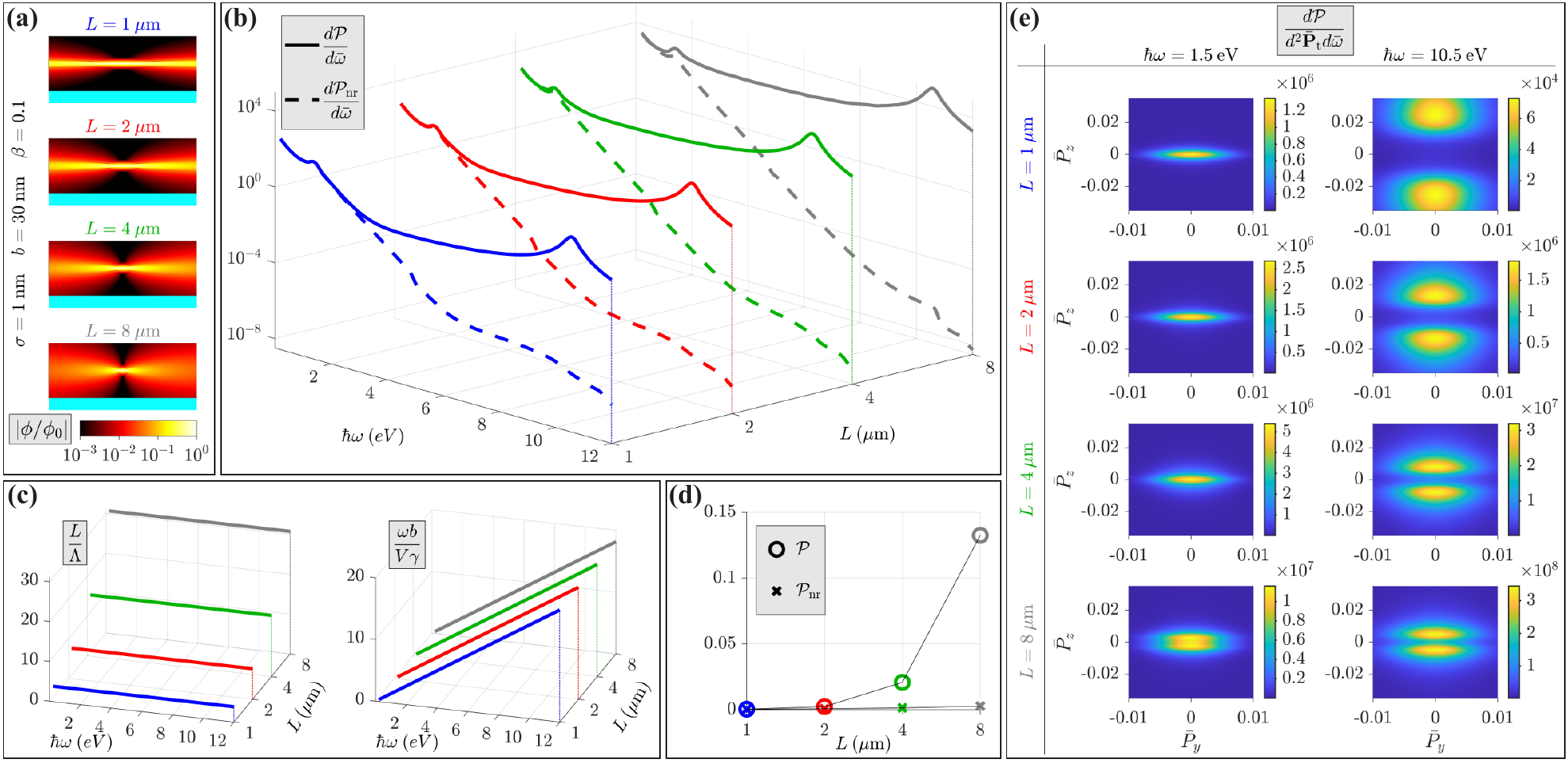}
\caption{Interaction length dependence of electron inelastic scattering. (a) Impinging electron beam and Aluminum slab mutual geometries pertaining the four selected slab lengths ($|\phi_0|$ is the maximum of the absolute value $|\phi|$). (b) Energy loss probability evaluated by using our approach $d {\mathcal P}/d\bar \omega$  (solid lines) and the no-recoil one $d{\mathcal P}_{\rm nr}/d\bar \omega$ (dashed lines) ($\bar \omega  = \hbar \omega / ( {mc^2 }  ))$. Their mutual discrepacy signals the impact of momentum recoil on the scattering process and it is more significant for longer slabs and higher energy losses. (c) Fundamental ratios identifying the quantum-recoil regime through Eqs.(\ref{Ineq1}) and (\ref{Ineq2}) and confirming the recoil phenomenology reported in subplot(b). (d) Comparison of total energy loss probabilty $\mathcal P$ and its no-recoil counterpart ${\mathcal P}_{\rm nr}$. (e) Momentum resolved energy loss probability at the Aluminium interband transition peak (first column) and surface plasmon peak (second column), where momentum recoil is negligible and relevant, respectively (${\bf{\bar P}}_{\rm t}  = {\bf{P}}_{\rm t} /P_0$). } 
\label{Fig2}
\end{figure*}

\begin{figure*}
\centering
\includegraphics[width = 1\linewidth]{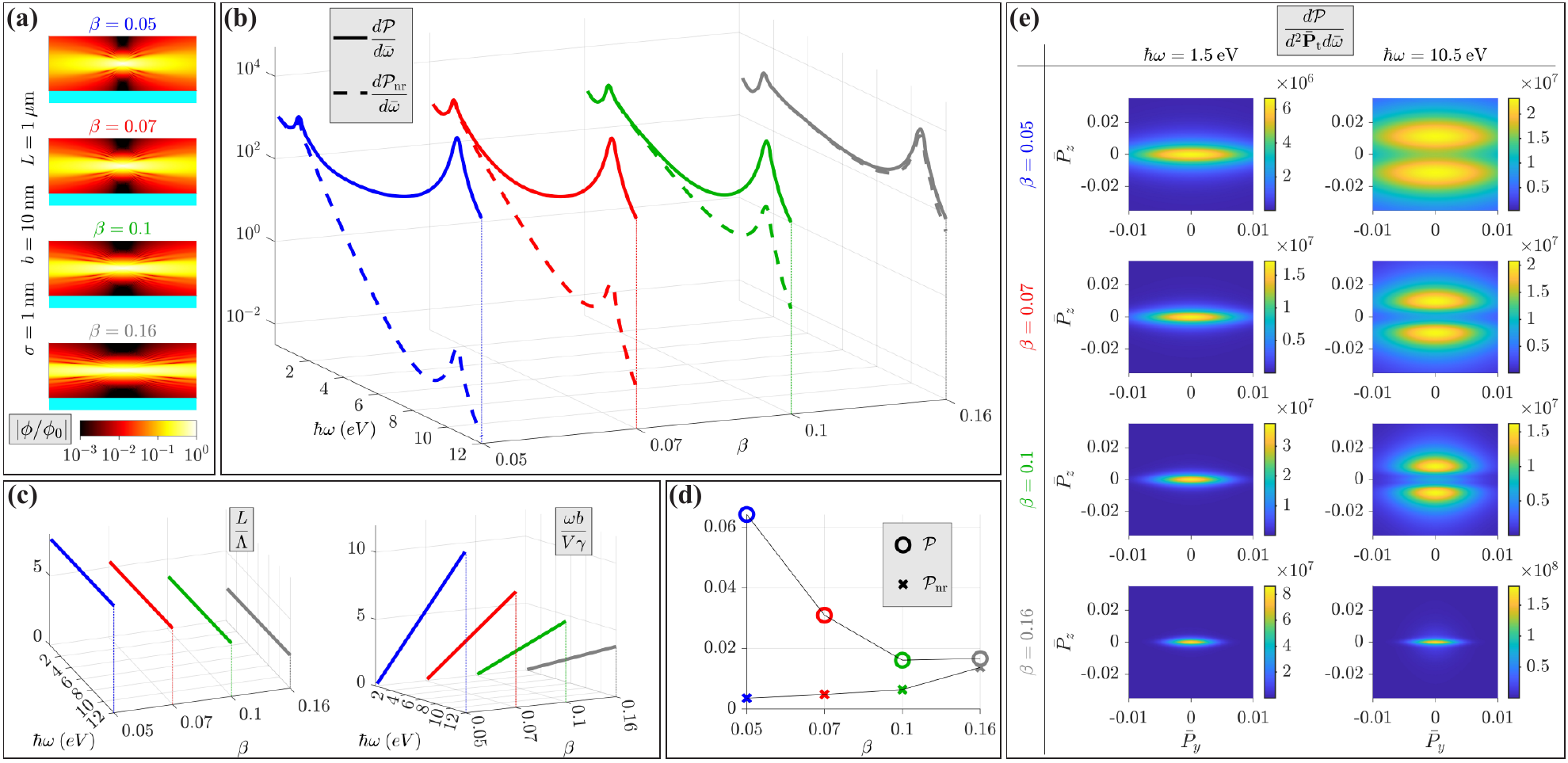}
\caption{Kinetic energy (electron velocity) dependence of electron inelastic scattering. See Fig.2 for definition details. (a) Impinging electron beam and Aluminum slab mutual geometries pertaining the four selected electron velocities. (b) The mutual discrepacy of $d {\mathcal P}/d\bar \omega$ and $d{\mathcal P}_{\rm nr}/d\bar \omega$ shows that the impact of momentum recoil on the scattering process (and related quantum features) is stronger at lower electron velocities and higher energy losses. (c) Fundamental ratios confirming the recoil phenomenology repored in subplot (b). (d) Total energy loss probability $\mathcal P$ and its no-recoil counterpart ${\mathcal P}_{\rm nr}$. (e) Momentum resolved energy loss probability.} 
\label{Fig3}
\end{figure*}

\section{Quantum-recoil regime}
In order to identify the actual regime where  significant quantum features show up in the ELP owing to momentum recoil (i.e. where $d {\mathcal P} /d\omega$ of Eq.(\ref{EnLoPr}) substantially differs from $d {\mathcal P}_{\rm nr} /d\omega$ of Eq.(\ref{EnLoPrCl})), consider the paradigmatic setup of Fig.1(a) where the sample is a slab of length $L$ and $\phi$ is the envelope of a paraxial Gaussian electron beam of main momentum ${\bf{P}}_0$ along the $x$-axis,  waist radius $w(0)= \sigma \gg \hbar /P_0$, and impact parameter $b$. As detailed in Appendix B, the corresponding broadening length of the beam is
\begin{equation} \label{LAMBDA}
\Lambda  = \beta \gamma \frac{{\sigma ^2 }}{{2\lambda _c }},
\end{equation}
where $\beta  = V/c$ and $\lambda _c  = \hbar /\left( {mc} \right) \simeq 3.86 \cdot 10^{-4} \: {\rm nm}$ is the reduced Compton wavelength of the electron. In Fig.1(b) we plot $\Lambda(\sigma)$ for various $\beta$ (and corresponding kinetic energies $K = mc^2 \left( {\gamma  - 1} \right)$) showing that the broadening length of nanometer-sized electron beams is generally in the micron scale or larger at relativistic kinetic energies whereas it is substantially smaller in the sub-relativistic regime $K  \lesssim 10 \: {\rm keV}$ and it rapidly decreases at lower kinetic energies. As a conseguence, for nanometer-sized fast electron beams probing nano-samples, the condition $L \ll \Lambda$ is markedly satisfied so that, as plotted in Fig.1(c1), the envelope $\phi$ is nearly undistorted over the interaction length with an almost uniform phase profile, this justifying the classical description of the interaction together with the accuracy of $d {\mathcal P}_{\rm nr} /d\omega$ of Eq.(\ref{EnLoPrCl}). Besides, the detection of a slab optical excitation is performed by maximizing the spatial overlap between its vacuum evanescent profile $\sim {\rm exp} (-z/d)$ and the electron beam envelope, i.e.  by setting $b \approx  d$ (see Fig.1(c1)) which leads to the necessary condition $\omega b/(V\gamma ) \approx 1$, since the vacuum decay length of an optical excitation of frequency $\omega$ created by the  electron is $d = V\gamma /\omega$.

In this paper we focus on quite the opposite regime $L > \Lambda$ which, as shown in Fig.1(b), can be accessed by resorting to sub-relativistic nanometer-sized electron beams interacting with micron-sized (mesoscopic) sample. The typical situation, whose geometry is sketched in Fig.1(c2), does not admit a classical description primarly since the spatial broadening of the envelope $\phi$ is genuinely a quantum effect which is taken into account by $d {\mathcal P} /d\omega$ of Eq.(\ref{EnLoPr}). Accordingly quantum features get more significant in the ELP as the ratio $L/\Lambda$ increases, i.e. as the electron kinetic energy decreases and the interaction length increases. In addition, as discussed in Section II (and Appendix B), quantum broadening is accompanied by a non-uniform phase profile of the envelope $\phi$ which consequently displays spatial oscillations becoming very rapid along its lateral sides. As a consequence, when detecting a sample optical excitation of frequency $\omega$ and transverse decay length $d = V \gamma /\omega$ (see Fig.1(c2)), quantum features become more significant as the impact parameter is increased with $b \gg d$ since the above lateral oscillations get stronger along the excitation near-field profile. This leads to the condition $\omega b/ (V\gamma) \gg 1$ which shows, in particular, that ELP quantum features more evidently show up when detecting optical excitations of higher frequencies.

\section{Inelastic scattering of aloof electrons by a metallic slab}
We now quantitatively examine the above discussed quantum-recoil regime in the specific and experimentally feasible scattering setup sketched in Fig.1(a). The aloof electron beam propagates in vacuum parallel to the $x$-axis (i.e. ${\bf{P}}_0  = P_0 {\bf{e}}_x$) and it is inelastically scattered by a metallic slab lying in the region $z<0$ and of finite length $L$ along the $x$-axis. With the aim of analytically studying the onset of quantum-recoil features, we assume that the initial electron envelope has the Gaussian-exponential profile $\phi \left( {{\bf{R}}_{\rm t} } \right) = \left[ {2/\left( {\pi \sigma ^4 } \right)} \right]^{1/4} \exp \left( { - Y^2 /\sigma ^2 } \right)\exp \left( { - \left| {Z - b} \right|/\sigma } \right)$ of width $\sigma \gg \hbar /P_0$ and impact parameter $b>0$. Moreover, we model the finite slab Green tensor as the product of the infinite slab Green tensor and the truncation function $\exp \left( { - 2\left| X \right|/L} \right)$. With these assumptions, the MELP of Eq.(\ref{MoEnLoPr}) can be casted as a single integral over the parallel wavevector ${\bf{k}}_\parallel   = k_x {\bf{e}}_x  + k_y {\bf{e}}_y$ labelling the optical excitations of the slab (see Appendix D), that is
\begin{equation} \label{MoReEnLoPrSLAB}
\frac{{d{\cal P}}}{{d^2 {\bf{P}}_{\rm t} d\omega }} = \frac{{4\alpha \sigma ^2 }}{{c\hbar ^2 \sqrt {2\pi ^3 } }}{\mathop{\rm Re}\nolimits}\left[ \int {d^2 {\bf{k}}_\parallel  } \Upsilon e^{ - \frac{{\sigma ^2 }}{{2\hbar ^2 }}\left( {P_y  - \hbar k_y } \right)^2 } J^ +  J^ - \right], 
\end{equation}
where  $\Upsilon \left( {{\bf{k}}_\parallel  ,\omega } \right)$ is a factor containing the TE and TM  reflection coefficients of the slab (see Eq.(\ref{Upsil})) and we have set
\begin{eqnarray} \label{Jtau}
J^\tau   = F_0^\tau  \;e^{i\left( {\tau \frac{{P_z }}{\hbar } + \sqrt {\frac{{\omega ^2 }}{{c^2 }} - k_\parallel ^2 } } \right)b}  + F_\sigma ^\tau  \;e^{ - \frac{b}{\sigma }}  + F_ + ^\tau  e^{\frac{i}{\hbar }q_ +  b}  + F_ - ^\tau  e^{\frac{i}{\hbar }q_ -  b} \nonumber \\
\end{eqnarray}
where $\tau = \pm 1$, $F_a ^\tau  \left( {{\bf{P}}_{\rm t} ,\omega \left| {{\bf{k}}_\parallel  } \right.} \right)$ are non-exponential functions (see Eqs.(\ref{F})) and 
\begin{equation} \label{qpm}
q_ \pm   = \sqrt {2\hbar P_0 \left[ {\left( {k_x  + \frac{\omega }{V}} \right) \pm \frac{{2i}}{L}} \right] + P_{\rm t}^2  - \left( {P_y  - \hbar k_y } \right)^2 }.
\end{equation}
with ${\mathop{\rm Im}\nolimits} \left( {q_ \pm  } \right) > 0$. Equation (\ref{MoReEnLoPrSLAB}) states that the contribution of the optical excitation of wavevector ${\bf k}_\parallel$ to the integral is proportional to $J^+ J^-$ and Eq.(\ref{Jtau}) analitically captures its full dependence on the impact parameter $b$, by means of four terms. The first term ($0$) accounts for the excitation vacuum profile (since $\sqrt{(\omega/c)^2 - k^2_\parallel}$ is the wavevector orthogonal to the slab), the second term ($\sigma$) describes the effect of the beam transverse confinement while the third and the fourth terms ($+$ and $-$) characterize the impact of the finite length of the slab (since $q_\pm$ depend on $L$). Now the remarkable point is that, in the relativistic limit $P_0 \rightarrow +\infty$ (where the effect of momentum recoil disappears, see the first of Eq.(\ref{HeHint})), the ($\pm$) contributions in Eq.(\ref{Jtau}) vanish since ${\mathop{\rm Im}\nolimits} \left( {q_ \pm  } \right) \to  + \infty$ whereas the ($0$) and ($\sigma$) contributions survive. We conclude that the last two ($\pm$) terms in Eq.(\ref{Jtau}) account for the effect of momentum recoil on inelastic scattering an therefore they quantitatively describe the discrepancy between our treatment and the no-recoil approach. Such interpretation of the ($\pm$) terms is corroborated by the fact that the exponentials $\exp \left( {iq_ \pm  b/\hbar } \right)$ vanish for $L \rightarrow 0$ (since ${\mathop{\rm Im}\nolimits} \left( {q_ \pm  } \right) \to  + \infty$) whereas they become unimodular in the limit $L \rightarrow \infty$ (since ${\mathop{\rm Im}\nolimits} \left( {q_ \pm  } \right) \to  0$), asymptotic behaviors in full agreement with the fact that momentum recoil is effective only if 
\begin{equation} \label{Ineq1}
\frac{L}{\Lambda} > 1,
\end{equation}
as discussed in Sections II and III. Moreover, such interpretation enables to find the conditions under which the quantum features of the interaction significantly show up in the detection of an optical excitation as a consequence of momentum recoil. To this end note that the function $F_0 ^\tau$ is practically not vanishing only for $k_x \simeq -\omega/V$ (see the first of Eqs.(\ref{F})) and consequently the first contribution in Eq.(\ref{Jtau}) is proportional to the well known decaying factor $\exp \left( { - \omega b/( V\gamma)} \right)$. Therefore if the conditions
\begin{equation} \label{Ineq2}
 \frac{{\omega b}}{{V \gamma}} \gg 1
\end{equation}
and $b \gg \sigma$ are met, the first two terms 
($0$  and $\sigma$) in Eq.(\ref{Jtau}) are negligible with respect to the the last two ($\pm$) terms which means that the effect of momentum recoil is dominant. Note that the condition in Eq.(\ref{Ineq2}) coincides with the one deduced in the qualititive discussion at the end of Section III where the enhancement of the quantum features in the detection of the excitation where ascribed to the rapid oscillations accompanying the broadening of the beam.

\begin{figure*}
\centering
\includegraphics[width = 1\linewidth]{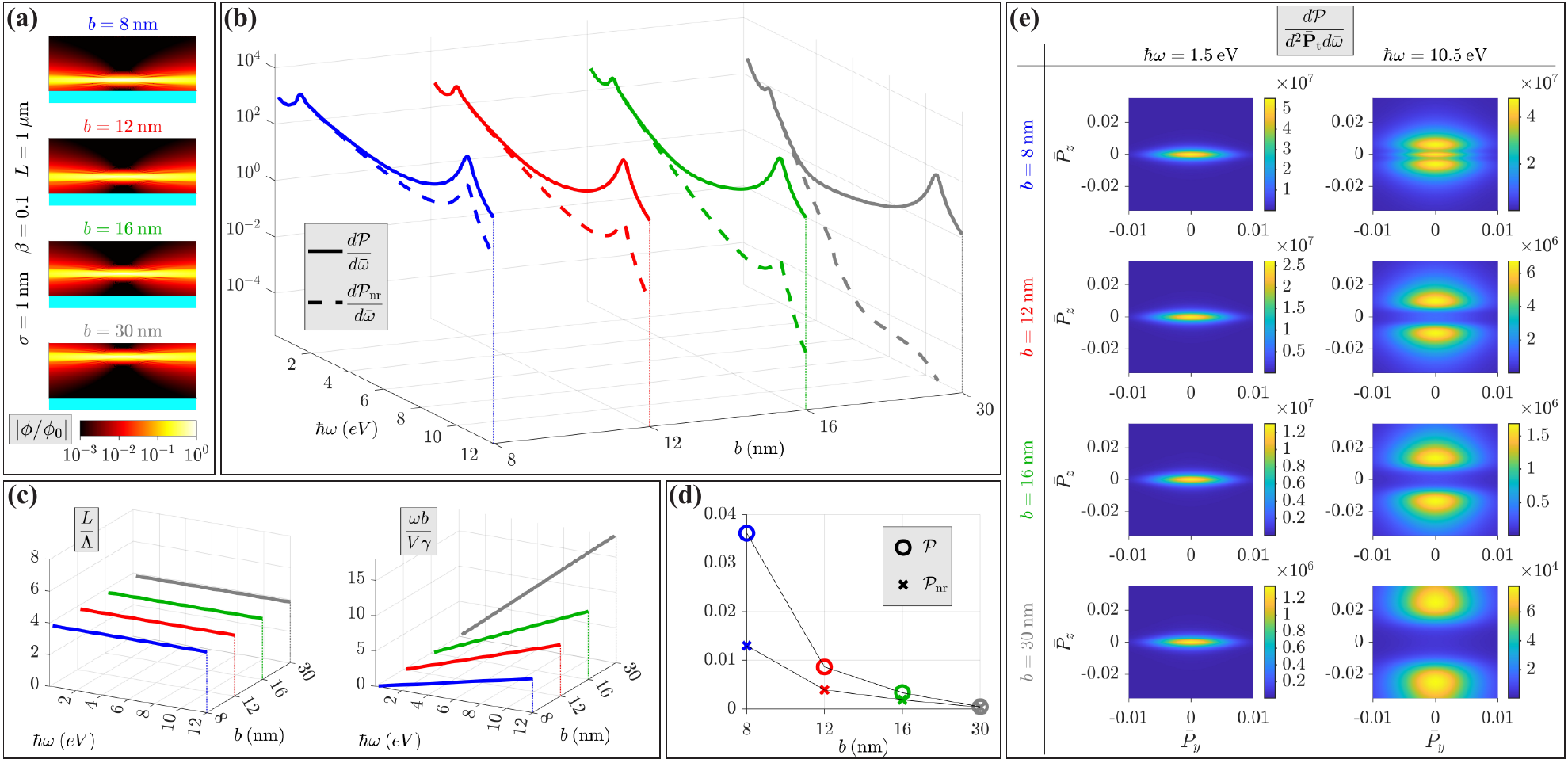}
\caption{Impact parameter dependence of electron inelastic scattering. See Fig.2 for definition details. (a) Impinging electron beam and Aluminum slab mutual geometries pertaining the four selected impact parameters. (b) The mutual discrepacy of $d {\mathcal P}/d\bar \omega$ and $d{\mathcal P}_{\rm nr}/d\bar \omega$ signals larger impact of momentum recoil on the scattering process at larger impact parametes and higher energy losses. (c) Fundamental ratios confirming the recoil phenomenology repored in subplot (b). (d) Total energy loss probability $\mathcal P$ and its no-recoil counterpart ${\mathcal P}_{\rm nr}$. (e) Momentum resolved energy loss probability.} 
\label{Fig4}
\end{figure*}

In order to numerically investigate the quantum-recoil regime, we have chosen aluminum as the metal of the slab  and we have used Eq.(\ref{MoReEnLoPrSLAB}) to evaluate both the ELP $d {\mathcal P}/d  \omega$ (after integration over ${\bf P}_{\rm t}$) and  its no-recoil counterpart $d{\mathcal P}_{\rm nr}/d  \omega$ (after taking the formal limit $P_0 \rightarrow +\infty$) in a number of electron-slab mutual configurations and we have reported the results of our simulations in Figs.2, 3 and 4. Aluminum is particularly suitable for our analysis since it displays marked spectral features at both  relatively low and high frequencies, with an interband transition peak at $1.2 \, {\rm eV}$ and a surface plasmon peak at $10.6 \, {\rm eV}$ \cite{Gerar}. 

In Fig.2 we analyze the effect of increasing the 
interaction length on electron inelastic scattering, in the quantum-recoil regime. We consider an impinging electron beam with $\sigma = 1 \, {\rm nm}$, $b = 30 \, {\rm nm}$ and $\beta = 0.1$ and four scattering geometries with slab lengths $L = 1,2,4$ and $8 \, {\rm \mu m}$, as we have pictorially sketched in Fig.2(a) to highlight beam broadening and beam-slab mutual positionings. In Fig.2(b) we plot the corresponding profiles of $d {\mathcal P}/d\bar \omega$ (solid lines) and $d{\mathcal P}_{\rm nr}/d\bar \omega$ (dashed lines) (where $\bar \omega  = \hbar \omega / ( {mc^2 }  )$), for energy losses $\hbar \omega$ up to $12 \, {\rm eV}$. Their mutual discrepancy is dramatically evident and it is seen to rise as $L$ and/or $\hbar \omega$ increase so that, as expected from the above discussion, the impact of recoil gets more significant for longer slabs and larger energy losses. This is in full agreement with the fact that the corresponding ratios $L/\Lambda$ and $\omega b /(V \gamma)$, plotted in Fig.2(c), satisfy the inequalities of Eqs.(\ref{Ineq1}) and (\ref{Ineq2}) for most of the considered energy losses. The only exception is for the range $\hbar \omega < 2 \, {\rm eV}$ (comprising the Aluminium interband transition peak) where the ratio $\omega b /(V \gamma)$ can get smaller than one so that recoil is negligible even though $L/\Lambda$ is always larger than one, as confirmed by the agreement of $d {\mathcal P}/d\bar \omega$ and $d{\mathcal P}_{\rm nr}/d\bar \omega$  over this range of energy losses in Fig.2(b). At higher energy losses, $d{\mathcal P}_{\rm nr}/d\bar \omega$ is generally very small as a consequence of the relatively large impact parameter (to the point that the surface plasmon peak does not actually show up in its profile) whereas $d {\mathcal P}/d\bar \omega$ turns out to be several order of magnitude larger. The global comparison between our approach and the no-recoil one is provided by Fig.2(d) where we plot the total ELP, ${\mathcal P} = \int {d\omega ({{d{\mathcal P}}}/{{d\omega }}})$, pertaining the four considered values of $L$ together with its no-recoil counterpart ${\mathcal P}_{\rm nr} = \int {d\omega ({{d{\mathcal P}_{\rm nr}}}/{{d\omega }}})$. Note that the total detected signal evidently increases with the interaction length, an important experimental benefit which can be properly described only by considering the effect of electron recoil. Moreover, the large value of $\mathcal P$ pertaining the longest slab ($L= 8 \: {\rm \mu m}$) supports the observation at the end of Section II about the possibility of accessing the electron-field strong coupling regime (which is beyond the approach discussed in the present paper) by resorting to even longer slabs. In Fig.2(e) we plot the MELP $d {\mathcal P} / {d^2 \bar{\bf{P}}_{\rm t} d\bar\omega }$ (where ${\bf{\bar P}}_{\rm t}  = {\bf{P}}_{\rm t} /P_0$) corresponding to the four considered slab lengths (rows), at the two energy losses $\hbar \omega = 1.5 \, {\rm eV}$ (left column) and $\hbar \omega = 10.5 \, {\rm eV}$ (right column), where momentum recoil is negligible and  significant, respectively (as discussed above). While in the left column the MELP shape is almost unaffected by the slab length and its peak value increases by $\sim 10$ from the top to the bottom, in the right column the situation is very different and it reveals a more involved evolution dictated by momentum recoil. In fact, momentum exchange with the slab splits the no-recoil peak into two lobes whose mutual separation decreases and whose peak values increase by $\sim 10^4$ as the slab length is increased, from the top to the bottom of the right column. In addition to shading light on momentum exchange dynamics, the results contained in Fig.2(e) provide evidence that, in the considered situations, electron-slab interaction does not invalidate the paraxial condition $| {\bf P} - {\bf{P}}_0  | \ll P_0$ and does not trigger nonlocal spatial effects since the maximum exchanged parallel wavevector (see Eq.(\ref{Kj}))
\begin{equation}
K_\parallel   = \sqrt {\left( {\frac{\omega }{V} + \frac{{P_{\rm t}^2 }}{{2\hbar P_0 }}} \right)^2  + \frac{{P_y^2 }}{{\hbar ^2 }}} 
\end{equation}
is found to be always much smaller than the Aluminum Fermi wavenumber $k_{\rm F} \simeq 1.75 \: \AA ^{-1}$ \cite{Krane}.
 
In Fig.3 we examine the effect of decreasing the electron velocity (kinetic energy) on electron inelastic scattering, in the quantum-recoil regime. We consider a slab of length $L = 1 \, {\rm \mu m}$ which separately scatters four impinging electron beams with $\sigma = 1 \, {\rm nm}$, $b = 10 \, {\rm nm}$ and $\beta = 0.05, 0.07, 0.1$ and $0.16$. Subplot pattern of Fig.3 is identical to that of Fig.2 so that we here omit repeating definition details. Figure 3(a) visually highlights the impact of electron velocity on beam broadening and it makes it clear that the envelope is always sufficiently separated from the slab. In Fig.3(b) the discrepacy of $d {\mathcal P}/d\bar \omega$ and $d{\mathcal P}_{\rm nr}/d\bar \omega$ clearly indicates that momentum recoil (and the quantum trait of the interaction) gets more relevant at lower electron velocities and again at higher energy losses, as expected and as confirmed by the fundamental ratios plotted in Fig.3(c). Remarkably, Fig.3(b) reveals that, in the considered range of energy loss, recoil is effectively negligible for $\beta > 0.16$, our predictions practically coinciding with those of the no-recoil approach at those electron velocities. The relevance of recoil in the sub-relativistic regime $\beta <0.16$ is also peculiarly highlighted by the discrepancy between $\mathcal{P}$ (the total ELP) and its no-recoil counterpart, as plotted in Fig.3(d). This is further confirmed by the MELP profiles plotted in Fig.3(e) at the four considered electron velocities (rows), since the two lobes whose splitting is produced by momentum recoil in the right column gradually merge into the no-recoil bell-shaped profile, from the top to the bottom. Conversely the MELP profiles in the left column are all bell shaped since the chosen energy loss is so small to effectively forbid recoil at the considered velocities (again the ratio $\omega b /(V \gamma)$ in Fig.3(c) can get very small for $\hbar \omega < 2 \, {\rm eV}$). Angain the MELP profiles in Fig.3(e) self-consistently validate the paraxial condition and ensure that nonlocal contributions to the slab response are negligible.

In Fig.4 we focus on the effect of increasing the impact parameter on electron inelastic scattering, in the quantum-recoil regime. We consider a slab of length $L = 1 \, {\rm \mu m}$ which separately scatters four impinging electron beams with $\sigma = 1 \, {\rm nm}$, $\beta = 0.1 $ and $b = 8, 12, 16$ and $30 \, {\rm nm}$. Subplot pattern of Fig.4 is identical to that of Figs.2 and 3 so that we again omit repeating definition details. Figure 4(a) displays impinging electron beam and Aluminium slab mutual geometries  pertaining the four considered impact parameters. The major result contained in Fig.4(b) is that recoil becomes more significant as the beam is moved away from the slab, since the discrepancy between $d {\mathcal P}/d\bar \omega$ and $d{\mathcal P}_{\rm nr}/d\bar \omega$ gets stronger as the impact parameter increases. Note that both $d {\mathcal P}/d\bar \omega$ and $d{\mathcal P}_{\rm nr}/d\bar \omega$ decrease as $b$ increases but the fall-off of the latter is dramatically more rapid to the point that at $b = 30 \, {\rm nm}$ the latter is six order of magnitude smaller than the former, in the range of energy loss comprising the surface plasmon peak. The results reported in subplots (c),(d) and (e) of Fig.4 are in agreement with the recoil phenomenology displayed in Fig.4(b).

\section{Conclusion}
In summary, we have investigated inelastic scattering of sub-relativistic (few keV) nanometer-sized electron beams by mesoscopic (micron-sized) samples in the aloof configuration. The considered combination of electron kinetic energies and interaction lengths entails a regime where electron recoil is significant and able to provide the scattering dynamics a marked quantum trait, as proved by the fact that the energy loss probability we have derived does not admit classical interpretation. Quantum features of inelastic scattering are found to get stronger as the electron kinetic energy is decreased and/or the interaction length is increased. Moreover, quantum broadening of the electron beam allows the energy loss probability to be significant at relatively large impact parameters and large energy losses, conditions where the classical prediction is practically vanishing. In addition to describe the impact of recoil on electron inelastic scattering, our results suggest that very large interaction lengths could in principle trigger the onset of an electron-matter strong-coupling regime which is structurally affected by quantum recoil. The description of such regime would require a nonperturbative refinement of our treatment which could possibly also account for the presence of field excitations before electron scattering, thus additionally enabling the investigation of photon-induced near-field electron microscopy (PINEM) in the quantum-recoil regime.

{\bf ACKNOWLEDGEMENTS} 

The author acknowledges PRIN 2017 PELM (grant number 20177PSCKT).

\appendix

\section{Paraxial Hamiltonian with momentum recoil}
The minimal coupling Hamiltonian of the electron in the presence of the quantized field of Eq.(\ref{Apot}) is $\hat H_{ei} = c\sqrt {m^2 c^2  + [ {{\hat{\bf P}} + e{\hat{\bf A}} ( {{\hat{\bf R}}}  )} ]^2 }$ where $m$ and $-e < 0$ are the electron rest mass and charge, and $\hat{\bf R}$ and $\hat{\bf P}$ are the electron position and momentum operators. To simplify the Hamiltonian, we now perform the paraxial approximation 
which is adequate for electron states whose momentum distribution is strongly peaked around a main momentum ${\bf{P}}_0 = \gamma m{\bf{V}} $ (i.e. $| {{\hat{\bf P}} - {\bf{P}}_0 } | \ll P_0$) where ${\bf{V}} = c{\bf{P}}_0 /\sqrt {m^2 c^2  + P_0^2 }$ is the main electron velocity and $\gamma  = 1/\sqrt {1 - (V/c) ^2 }$. Owing to paraxiality and to the weakness of the electron-field coupling, the Hamiltonian $\hat H_{ei}$ slightly departs from the main energy $\gamma mc^2$ and the first order correction is $\hat H_{ei}  - \gamma mc^2  \simeq \hat H_e  + \hat H_{{\mathop{\rm int}} }$ where 
\begin{eqnarray} \label{APPHeHint}
\hat H_e  &=& \frac{V}{2P_0}( {\hat P^2  - P_0^2 } ), \nonumber \\
\hat H_{{\mathop{\rm int}} } & =& e{\bf{V}} \cdot {\hat{\bf A}}
\end{eqnarray}
are the free electron and interaction Hamiltonians. In deriving these expressions, we have used the Coulomb gauge relation ${\bf{\hat A}} ( {{\bf{\hat R}}}  ) \cdot {\bf{\hat P}} = {\bf{\hat P}} \cdot {\bf{\hat A}} ( {{\bf{\hat R}}}  )$, we have set $ {\hat{\bf P}} \cdot {\hat{\bf A}} \simeq {\bf{P}}_0  \cdot {\hat{\bf A}}$ in the interaction Hamiltonian and we have neglected the term $e^2 \hat A^2$. After decomposing the momentum operator into its parts ${\hat{\bf P}}_{\rm t}$ transverse and $\hat P_{\rm v}  {\bf{u}}_{\rm v}$ parallel to the velocity unit vector ${\bf{u}}_{\rm v}  = {\bf{V}}/V$, the evident recasting of the electron Hamiltonian
\begin{equation} \label{Hefull}
\frac{{\hat H_e }}{{VP_0 }} = \frac{{( {\hat P_{\rm v}  - P_0 } )}}{{P_0 }} + \frac{{\hat P_{\rm t}^2  + ( {\hat P_{\rm v}  - P_0 } )^2 }}{{2P_0^2 }}
\end{equation}
elucidates that $V( {\hat P_{\rm v}   - P_0 } )$ is its leading contribution due to paraxiality $ | {{\hat{\bf P}}_{\rm t} }  | \ll P_0$, $ | {\hat P_{\rm v}  - P_0 }  | \ll P_0$. The small quadratic terms account for momentum recoil whose main effect is to reshape the electron wave function so that they can be safely neglected over ultra-fast interaction times (no-recoil approximation). On the other hand, in this paper we are considering scattering setups where the larger interaction time can in principle magnify the role of momentum recoil so that we retain the quadratic term ${\hat P_{\rm t}^2 }$ while we discard the quadratic term ${( {\hat P_{\rm v}  - P_0 } )^2 }$. The fact that longitudinal momentum recoil provides a much weaker effect than the transverse one is easily justified since Eq.(\ref{Hefull}) can be casted as
\begin{equation}
 {\frac{({\hat P_{\rm v}  - P_0 })}{{P_0 }}}  =  - 1 + \sqrt {1 +    {\frac{{2\hat H_e }}{{VP_0 }} - \frac{{\hat P_{\rm t}^2 }}{{P_0^2 }}}  }  \simeq \frac{{\hat H_e }}{{VP_0 }} - \frac{{\hat P_{\rm t}^2 }}{{2P_0^2 }},
\end{equation}
where we have noted that $|\hat H_e | \ll VP_0$ in paraxial approximation, whose comparison with Eq.(\ref{Hefull}) reveals that ${( {\hat P_{\rm v}  - P_0 } )^2 }$ can be self-consistently neglected. Therefore, after dropping the irrelevant constant contributions and using Eq.(\ref{Apot}), Eqs.(\ref{APPHeHint}) turn into 
\begin{eqnarray} 
\hat H_e  &=& V\left( {\hat P_{\rm v}  + \frac{{\hat P_{\rm t}^2 }}{{2P_0 }}} \right), \nonumber \\ 
 \hat H_{{\mathop{\rm int}} }  &=& \int {d\xi } \left[ {C(  {\hat{\bf{R}},\xi } )\hat f\left( \xi  \right) + C^* ( {\hat{\bf{R}},\xi } )\hat f^\dag  \left( \xi  \right)} \right],  
\end{eqnarray}
where $C\left( {{\bf{R}},\xi } \right) = \hbar \omega  \sqrt {\left( {4\alpha /c^3 } \right){\rm Im}\;\varepsilon _\omega  \left( {\bf{r}} \right)} \; {\bf V } \cdot {\mathcal G}_\omega  \left( {{\bf{R}},{\bf{r}}} \right){\bf{e}}_j$ and $\alpha  = e^2 /\left( {4\pi \varepsilon _0 \hbar c} \right) \simeq 1/137$ is the fine-structure constant.

\section{Incident electron states}
Since electron beams routinely used in scattering experiments are nearly monoenergetic, the incident electron states $\left| {\psi _0 } \right\rangle$ are nearly eigenstates of the free electron Hamiltonian $\hat H_e$ in the first of Eqs.(\ref{HeHint}), meaning that their energy distribution is strongly peaked around a reference value $E_0$. Both to analyze the structure of such incident electron states and to subsequently describe their scattering, it is convenient to choose the representation induced by the eigenvectors of $\hat H_e$. We hereafter label with $\left| {\bf{R}} \right\rangle$ and $\left| {\bf{P}} \right\rangle$ the eigenvectors of the position ${{\bf{\hat R}}}$ and momentum ${{\bf{\hat P}}}$ operators, respectively, satisfying  $\left\langle {{\bf{R}}} \mathrel{\left | {\vphantom {{\bf{R}} {\bf{P}}}} \right. \kern-\nulldelimiterspace}  {{\bf{P}}} \right\rangle  = e^{\frac{i}{\hbar }{\bf{P}} \cdot {\bf{R}}} /\left( {2\pi \hbar } \right)^{3/2}$ and the usual orthonormalization ${\left\langle {{\bf{R}}} \mathrel{\left | {\vphantom {{\bf{R}} {{\bf{R}}'}}} \right. \kern-\nulldelimiterspace}
 {{{\bf{R}}'}} \right\rangle  = \delta \left( {{\bf{R}} - {\bf{R}}'} \right)}$, ${\left\langle {{\bf{P}}}\mathrel{\left | {\vphantom {{\bf{P}} {{\bf{P}}'}}} \right. \kern-\nulldelimiterspace}  {{{\bf{P}}'}} \right\rangle  = \delta \left( {{\bf{P}} - {\bf{P}}'} \right)}$, and closure  ${\int {d^3 {\bf{R}}} \left| {\bf{R}} \right\rangle \left\langle {\bf{R}} \right| = \hat I}$, ${\int {d^3 {\bf{P}}} \left| {\bf{P}} \right\rangle \left\langle {\bf{P}} \right| = \hat I}$ relations. 
 
Since each of the three components of the momentum operator commutes with the free electron Hamiltonian in the first of Eqs.(\ref{HeHint}), ${\bf{\hat P}}_{\rm t}$ and $\hat H_e$ constitute a complete set of commuting observables and, denoting by ${\bf{P}}_{\rm t}$ and $E$ the respective eigenvalues, their common eigenvectors  are easily seen to be
\begin{equation} \label{PtE}
\left| {{\bf{P}}_{\rm t} ,E} \right\rangle  = \frac{1}{{\sqrt V }}\left| {\bf{P}} \right\rangle _{{\bf{P}} = {\bf{P}}_{\rm t}  + \left( {\frac{E}{V} - \frac{{P_{\rm t}^2 }}{{2P_0 }}} \right){\bf{u}}_{\rm v} } 
\end{equation}
which satisfy the orthonormalization and closure relations
\begin{eqnarray} \label{OnClel}
&& \left\langle {{{\bf{P}}_{\rm t} ,E}}
 \mathrel{\left | {\vphantom {{{\bf{P}}_{\rm t} ,E} {{\bf{P}}'_{\rm t} ,E'}}}
 \right. \kern-\nulldelimiterspace}
 {{{\bf{P}}'_{\rm t} ,E'}} \right\rangle  = \delta \left( {{\bf{P}}_{\rm t}  - {\bf{P}}'_{\rm t} } \right)\delta \left( {E - E'} \right), \nonumber \\
&& \int {d^2 {\bf{P}}_{\rm t} } \int {dE} \left| {{\bf{P}}_{\rm t} ,E} \right\rangle \left\langle {{\bf{P}}_{\rm t} ,E} \right| = \hat I.
 \end{eqnarray}
In plain English, the eigenvector of $\hat H_e$ of transvere momentum ${{\bf{P}}_{\rm t} }$ and energy $E$ is (up to a normalization factor) the momentum eigevenctor with the same transverse component and longitudinal component $P_{\rm v}  = E/V - P_{\rm t}^2 /\left( {2P_0 } \right)$. 

The expansion of the incident state $\left| {\psi _0 } \right\rangle$ in terms of energy eigenvectors reads
\begin{equation} \label{initstate}
\left| {\psi _0 } \right\rangle  = \int {d^2 {\bf{P}}_{\rm t} } \int {dE} \;\bar \psi _0 \left( {{\bf{P}}_{\rm t} ,E} \right)\left| {{\bf{P}}_{\rm t} ,E} \right\rangle 
\end{equation}
where the spectral wave function $\bar \psi \left( {{\bf{P}}_{\rm t} ,E} \right)$ is normalized as $\int {d^2 {\bf{P}}_{\rm t} } \int {dE} \left| {\bar \psi_0 \left( {{\bf{P}}_{\rm t} ,E} \right)} \right|^2  = 1$. Evidently $\left| {\psi _0 } \right\rangle$ can not be an eigenstate of $\hat H_e$ since the corresponding spectral function $\bar \psi _0 \left( {{\bf{P}}_{\rm t} ,E} \right) = \bar \phi \left( {{\bf{P}}_{\rm t} } \right)\delta \left( {E - E_0 } \right)$ is not square-integrable. Therefore we consider normalized nearly monoenergetic incident states with energy uncertainty $\Delta E \ll E_0$ and, to ensure 
that the state is normalized also in the limit $\Delta E \to 0^ +$, we set
\begin{equation} \label{spect}
\bar \psi _0 \left( {{\bf{P}}_{\rm t} ,E} \right) = \bar \phi \left( {{\bf{P}}_{\rm t} } \right)\chi ^{\left(\Delta E\right)} \left( {E - E_0 } \right)
\end{equation}
where $\bar \phi \left( {{\bf{P}}_{\rm t} } \right)$ is normalized as $\int {d^2 {\bf{P}}_{\rm t} } \left| {\bar \phi \left( {{\bf{P}}_{\rm t} } \right)} \right|^2  = 1$ and 
\begin{equation}
\chi ^{\left( \eta  \right)} \left( \zeta  \right) = \frac{1}{{\sqrt {\pi \eta } }}{\rm sinc} \left( {\frac{\zeta }{\eta }} \right)
\end{equation}
satisfies the relations
\begin{eqnarray} \label{ChiDelta}
 \int\limits_{ - \infty }^{ + \infty } {d\zeta } \,\left[ {\chi ^{\left( \eta  \right)} \left( \zeta  \right)} \right]^2  &=& 1, \nonumber \\ 
\mathop {\lim }\limits_{\eta  \to 0^ +  } \left[ {\chi ^{\left( \eta  \right)} \left( \zeta  \right)} \right]^2  &=& \delta \left( \zeta  \right),
\end{eqnarray}
i.e. $\left[ {\chi ^{\left( \eta  \right)} \left( \zeta  \right)} \right]^2$ is a representation of the delta function. Since the momentum-space wave function of the incident state is easily seen to be 
\begin{equation}
\left\langle {{\bf{P}}}
 \mathrel{\left | {\vphantom {{\bf{P}} {\psi _0 }}}
 \right. \kern-\nulldelimiterspace}
 {{\psi _0 }} \right\rangle  = \bar \phi \left( {{\bf{P}}_{\rm t} } \right)\sqrt V \chi ^{\left( {\Delta E} \right)} \left[ {V\left( {P_{\rm v}  + \frac{{P_{\rm t}^2 }}{{2P_0 }}} \right) - E_0 } \right]
\end{equation}
we note that the necessary paraxial condition $| {{\bf P} - {\bf P}}_0  | \ll P_0$ is satisfied whenever $\bar \phi \left( {{\bf{P}}_{\rm t} } \right)$ is not negligible only for $\left| {{\bf{P}}_{\rm t} } \right| \ll P_0$ and $| {E_0  - VP_0 } | \ll VP_0$. A straightforward calculation shows that the wave function $\psi _0 \left( {\bf{R}} \right) = \left\langle {{\bf{R}}}  \mathrel{\left | {\vphantom {{\bf{R}} {\psi _0 }}}  \right. \kern-\nulldelimiterspace}  {{\psi _0 }} \right\rangle$ is given by
\begin{equation} \label{wafu}
\psi _0 \left( {\bf{R}} \right) = \theta \left( {\frac{\ell }{2} - \left| {R_{\rm v} } \right|} \right)\frac{1}{{\sqrt \ell  }}e^{\frac{{iE_0 }}{{\hbar V}}R_{\rm v} } \phi \left( {\bf{R}} \right)
\end{equation}
where $\theta \left( \xi  \right)$ is the Heaviside step function, $\ell  = 2\hbar V/\Delta E$, 
\begin{equation} \label{envelope}
\phi \left( {\bf{R}} \right) = \frac{1}{{2\pi \hbar }}\int {d^2 {\bf{P}}_{\rm t} } e^{\frac{i}{\hbar }\left( {{\bf{P}}_{\rm t}  \cdot {\bf{R}}_{\rm t}  - \frac{{P_{\rm t}^2 }}{{2P_0 }}R_{\rm v} } \right)} \bar \phi \left( {{\bf{P}}_{\rm t} } \right).
\end{equation}
and ${\bf{R}}_{\rm t}$ and $R_{\rm v} {\bf{u}}_{\rm v}$ are the transverse and longitudinal parts of the position vector $\bf R$ (see the inset of Fig.1(a)). Due to the presence of the $\theta$-function, $\psi _0 \left( {\bf{R}} \right)$ is not vanishing only in the region $ - \ell /2 < R_{\rm v}  < \ell /2$ so that the predictions based on the here chosen incident state $\left| {\psi _0 } \right\rangle$ coincides with those of the approach where quantization is performed in a box of length $\ell$. Accordingly the above discussed limit $\Delta E \to 0^ + $ amouts to the limit $\ell  \to  + \infty$ of the quantization box approach. 

The wave function $\psi _0 \left( {\bf{R}} \right)$ factorizes into a plane wave carrier of momentum $E_0 /V \simeq P_0$, propagating along the direction of the electron velocity, and a function $\phi \left( {\bf{R}} \right)$ which is a slowly varying envelope due to the above discussed paraxial requirement ($\bar \phi \left( {{\bf{P}}_{\rm t} } \right)$ is not negligible only for $\left| {{\bf{P}}_{\rm t} } \right| \ll P_0$). It is worth noting that structure of the wave function formally coincides with the one of paraxial electromagnetic fields \cite{Ciat2} and hence the incident electron states we are discussing exhibit the same spatial features of paraxial optical beams. Most notably for our purposes, paraxial optical diffraction corresponds to the spatial broadeing of the incident electron beam which is here due to the quadratic  term ${P_{\rm t}^2 }$ in Eq.(\ref{envelope}). To discuss this crucial point we start noting that a simple calculation shows that $\int {d^2 {\bf{R}}_{\rm t} } \left| {\phi \left( {\bf{R}} \right)} \right|^2  = 1$ for any $R_{\rm v}$ so that it is possible to introduce the trasverse average of an operator as $\langle {\hat O} \rangle _{R_{\rm v} }  = \int {d^2 {\bf{R}}_{\rm t} } \phi ^* \left( {\bf{R}} \right)\hat O\phi \left( {\bf{R}} \right)$ and accordingly define the transverse width of the envelope at the plane $R_{\rm v}$ as $\Delta R_{\rm t} \left( {R_{\rm v} } \right) = \sqrt { \langle {| {\hat {\bf{R}}_{\rm t}  -  \langle {\hat{\bf{R}}_{\rm t} }  \rangle _{R_{\rm v} } } |^2 }  \rangle _{R_{\rm v} } }$. Using Eq.(\ref{envelope}) to evaluate this average and assuming that the beam waist plane is at $R_{\rm v} = 0$, after some tedious but straightfoward algebra, we get the espressive and well known result $\Delta R_{\rm t} \left( {R_{\rm v} } \right) = \sqrt {w^2 \left( 0 \right) + \left( {{{\Delta P_{\rm t} }}/{{P_0 }}} \right)^2 R_{\rm v}^2 }$, which quantifies the broadening experienced by the electron beam. Here $\Delta P_{\rm t}  = \sqrt {\langle {| {{\bf{\hat P}}_{\rm t}  -  \langle {{\bf{\hat P}}_{\rm t} }  \rangle _0 } |^2 } \rangle _0 } \ll P_0$ is the transverse momentum uncertainty (which amounts to the width of the function $\bar \phi \left( {{\bf{P}}_{\rm t} } \right)$) so that the uncertainty relation $\Delta R_{\rm t} \left( 0 \right)\Delta P_{\rm t}  \ge \hbar /2$ sets the constraint 
\begin{equation}
\Delta R_{\rm t} \left( 0 \right) \gg \frac{\hbar}{2P_0}
\end{equation}
to the beam waist width and enables the estimation
\begin{equation} \label{width}
\Delta R_{\rm t} \left( {R_{\rm v} } \right) = \Delta R_{\rm t} \left( 0 \right)\sqrt {1 + \left( {\frac{{R_{\rm v} }}{\Lambda }} \right)^2 } 
\end{equation}
where the beam broadening length $\Lambda$ is given by
\begin{equation} \label{brlen}
\Lambda  = \frac{{P_0 }}{\hbar }\left[\Delta R_{\rm t} \left( 0 \right)\right]^2.
\end{equation}
The equivalent expression $\Lambda  = \beta \gamma \left[\Delta R_{\rm t} \left( 0 \right)\right]^2/\lambda _c$, where $\lambda _c  = \hbar /\left( {mc} \right) \simeq 3.86 \cdot 10^{-4} \: {\rm nm}$ is the reduced Compton wavelength of the electron, elucidates that the broadening length decreases as the electron velocity and/or the beam waist width are decreased. 

In order to examine the spatial features of the broadening beam it is convenient to derive an expression for $\phi \left( {\bf{R}} \right)$ not involving integration over the momentum space. To this end, after setting $R_{\rm v} =0$ in Eq.(\ref{envelope}) and inverting the Fourier integral, we get
\begin{equation} \label{phit}
\bar \phi \left( {{\bf{P}}_{\rm t} } \right) = \frac{1}{{2\pi \hbar }}\int {d^2 {\bf{R}}_{\rm t} } e^{ - \frac{i}{\hbar }{\bf{P}}_{\rm t}  \cdot {\bf{R}}_{\rm t} } \phi \left( {{\bf{R}}_{\rm t} } \right)
\end{equation}
which inserted back into (\ref{envelope}), after perfoming the Fresnel integrals, yields
\begin{equation} \label{envelopeR}
\phi \left( {\bf{R}} \right) = \int {d^2 {\bf{R}}'_{\rm t} } \; {\mathcal F} \left( {{\bf{R}}_{\rm t}  - {\bf{R}}'_{\rm t} \left| {R_{\rm v} } \right.} \right)\phi \left( {{\bf{R}}'_{\rm t} } \right),
\end{equation}
where 
\begin{equation} \label{APPFres}
{\mathcal F} \left( {{\bf{R}}_{\rm t} \left| {R_{\rm v} } \right.} \right) = \frac{1}{{i\pi }}\left( {\frac{{P_0 }}{{2\hbar R_{\rm v} }}} \right)e^{i\left( {\frac{{P_0 }}{{2\hbar R_{\rm v} }}} \right)R_{\rm t}^2 }
\end{equation}
is the Fresnel kernel. Equation (\ref{envelopeR}) shows that the envelope $\phi \left( {\bf{R}} \right)$ is fully determined by its waist profile $\phi \left( {{\bf{R}}_{\rm t} } \right)$ and consistency is ensured by the relation
\begin{equation}
\mathop {\lim }\limits_{R_{\rm v}  \to 0} F\left( {{\bf{R}}_{\rm t} \left| {R_{\rm v} } \right.} \right) = \delta \left( {{\bf{R}}_{\rm t} } \right),
\end{equation}
a direct consequence of the delta function representation $\mathop {\lim }\limits_{\eta  \to 0^ +  } \left( {1/\sqrt {i\pi \eta } } \right)e^{\frac{i}{\eta }\zeta ^2 }  = \delta \left( \zeta  \right)$. However, the Fresnel kernel in Eq.(\ref{envelopeR}) accurately behaves as the delta function whenever it has a very large number of oscillations over the spatial scale of $\phi \left( {{\bf{R}}'_{\rm t} } \right)$, i.e. whenever the condition $\hbar |R_{\rm v}| /P_0  \ll w^2 \left( 0 \right)$ is satysfied. Therefore, using Eq.(\ref{brlen}) we conclude that 
\begin{equation} \label{APPundist}
\phi \left( {\bf{R}} \right) \simeq \phi \left( {{\bf{R}}_{\rm t} } \right), \quad {\rm for} \; |R_{\rm v}| \ll \Lambda,
\end{equation}
or, in plain English, the envelope approximately does not undergo any distortion over longitudinal distances much smaller then its broadening length and the wave function $\psi _0 \left( {\bf{R}} \right)$ of Eq.(\ref{wafu}) provides a geometrical optics description of the incident electron beam. Another spatial phenomenon related to beam broadening, which is worth examining here, shows up at the sides of the beams. Without loss of generality we can assume that the beam axis is along the $R_{\rm v}$ axis (i.e. $\langle {{\bf{\hat R}}_{\rm t} } \rangle _0 = 0$) so that on the lateral sides of the beam the inequality $\left| {{\bf{R}}_{\rm t} } \right| \gg \Delta R_{\rm t} \left( 0 \right) \approx \left| {{\bf{R}}'_{\rm t} } \right|$ holds and the contribution of ${R^{'2}_{\rm t} }$ in the Fresnel kernel of Eq.(\ref{envelopeR}) can be neglected on the region where the beam broadens $|R_{\rm v}| > \Lambda$. As a consequence, after using Eq.(\ref{phit}) and Eq.(\ref{brlen}), we get the asymptotic expression
\begin{equation} \label{phsides}
\phi \left( {\bf{R}} \right) = \frac{{P_0 }}{{iR_{\rm v} }}e^{\frac{{i\Lambda }}{{2R_{\rm v} }}\left[ {\frac{{\left| {{\bf{R}}_{\rm t} } \right|}}{{\Delta R_{\rm t} \left( 0 \right)}}} \right]^2 } \bar \phi \left( {P_0 \frac{{{\bf{R}}_{\rm t} }}{{R_{\rm v} }}} \right)
\end{equation}
manifestly showing that the envelope $\phi \left( {\bf{R}} \right)$ displays lateral oscillations over the sides of the beam which get faster as $\left| {{\bf{R}}_{\rm t} } \right|$ increases.

The above discussed spatial features realted to beam broadening can be explicitly checked for Gaussian beams with envelope profile at the waist plane
\begin{equation} \label{Gauss}
\phi \left( {{\bf{R}}_{\rm t} } \right) = \sqrt {\frac{2}{{\pi \sigma ^2 }}} e^{ - \frac{{R_{\rm t}^2 }}{{\sigma ^2 }}},
\end{equation}
where $\sigma  = w\left( 0 \right) \gg \hbar /P_0 $ is the waist radius so that the waist transverse width is $\Delta R_{\rm t} \left( 0 \right) = \sigma /\sqrt 2$. Inserting Eq.(\ref{Gauss}) into Eq.(\ref{envelopeR}) and performing the Gaussian integral we get
\begin{equation} \label{phiGauss}
\phi \left( {\bf{R}} \right) = \frac{{\sqrt {\frac{2}{{\pi \sigma ^2 }}} }}{{\left( {1 + i\frac{{R_{\rm v} }}{\Lambda }} \right)}}e^{ - \frac{{\left( {1 - i\frac{{R_{\rm v} }}{\Lambda }} \right)}}{{\sigma ^2 \left[ {1 + \left( {\frac{{R_{\rm v} }}{\Lambda }} \right)^2 } \right]}}R_{\rm t}^2 } 
\end{equation}
where $\Lambda  = P_0 \sigma ^2 /\left( {2\hbar } \right)$. Evidently the beam broadens with its radius given by $w\left( {R_{\rm v} } \right) = \sigma \sqrt {1 + \left( {R_{\rm v} /\Lambda } \right)^2 }$ which agrees with Eqs.(\ref{width}) and (\ref{brlen}). In the region $R_{\rm v} \ll \Lambda$ the enevlope $\phi ({\bf R})$ very slightly departs from its waist profile $\phi ({\bf R}_{\rm t})$. In addition, for $R_{\rm v}  \gg \Lambda$, Eq.(\ref{phiGauss}) reduces to 
\begin{equation}
\phi \left( {\bf{R}} \right) \simeq \sqrt {\frac{2}{{\pi \sigma ^2 }}} \frac{\Lambda }{{iR_{\rm v} }}e^{i\frac{\Lambda }{{R_{\rm v} }}\left( {\frac{{R_{\rm t} }}{\sigma }} \right)^2 } 
\end{equation}
which agrees with Eq.(\ref{phsides}) and displays oscillations getting faster at the sides of the beam.

\section{Momentum resolved energy loss probability}
We label with $\left| 0 \right\rangle$ the vaccuum state of the field defined by $\hat f\left( \xi  \right)\left| 0 \right\rangle  = 0$ and $\left\langle {0}  \mathrel{\left | {\vphantom {0 0}}  \right. \kern-\nulldelimiterspace} {0} \right\rangle  = 1$. Due to the bosonic commutation relations $[ {\hat f\left( \xi  \right),\hat f\left( {\xi '} \right)} ] = 0$ and $[ {\hat f\left( \xi  \right),\hat f^\dag  \left( {\xi '} \right)} ] = \delta \left( {\xi  - \xi '} \right)$, where $\delta \left( {\xi  - \xi '} \right) = \delta \left( {{\bf{r}} - {\bf{r}}'} \right)\delta _{jj'} \delta \left( {\omega  - \omega '} \right)$, the kets $\left| {\xi _1  \ldots \xi _n } \right\rangle  =  (1 / \sqrt {n!} ) \prod\nolimits_{j = 1}^n {\hat f^\dag  \left( {\xi _j } \right)\left| 0 \right\rangle }$ (with $\left| {\xi _1  \ldots \xi _n } \right\rangle  = \left| 0 \right\rangle$ for $n=0$) are field states with $n$ quanta distributed on the $n$ polaritonic excitations ${\xi _1  \ldots \xi _n }$. Moreover, they are eigevenctors of the field Hamiltonian, $\hat H_{em} \left| {\xi _1  \ldots \xi _n } \right\rangle  = ( \sum\nolimits_{j = 0}^n {\hbar \omega _j } )\left| {\xi _1  \ldots \xi _n } \right\rangle$, and they satisfy the orthornormalization and closure relations
\begin{eqnarray} \label{OnClem}
&& \left\langle {{\xi _1  \ldots \xi _n }}
 \mathrel{\left | {\vphantom {{\xi _1  \ldots \xi _n } {\xi '_1  \ldots \xi '_{n'} }}}
 \right. \kern-\nulldelimiterspace}
 {{\xi '_1  \ldots \xi '_{n'} }} \right\rangle  = \frac{{\delta _{n,n'} }}{{n!}}\sum\limits_{\pi  \in S_n } {\left[ {\prod\nolimits_{j = 1}^n {\delta \left( {\xi _j  - \xi '_{\pi \left( j \right)} } \right)} } \right]}, \nonumber  \\ 
&& \sum\limits_{n = 0}^\infty  {\int {d\xi _1  \cdots d\xi _n } } \left| {\xi _1  \ldots \xi _n } \right\rangle \left\langle {\xi _1  \ldots \xi _n } \right| = \hat I,  
\end{eqnarray}
where the sum in the first equation spans the $n!$ permutation $\pi$ of the symmetric group $S_n$.  Therfore the kets $\left| {{\bf{P}}_{\rm t} ,E;\xi _1  \ldots \xi _n } \right\rangle  = \left| {{\bf{P}}_{\rm t} ,E} \right\rangle  \otimes \left| {\xi _1  \ldots \xi _n } \right\rangle$ are eigenstates of the unperturbed Hamiltonian $\hat H_0  = \hat H_e  + \hat H_{em}$ and, from Eqs. (\ref{OnClel}) and (\ref{OnClem}), they  are an ortonormal basis of the electron-field state space

To model the scattering of the electron initially prepared in the state $\left| {\psi _0 } \right\rangle$ of Eq.(\ref{initstate}), we choose the interaction representation and we set $\left| {\Psi \left( { - \infty } \right)} \right\rangle  = \left| {\psi _0 } \right\rangle  \otimes \left| 0 \right\rangle$ for the intial electron-field state (since the field is not excited at first). First order perturbation theory provides the state after the interaction $\left| {\Psi \left( { + \infty } \right)} \right\rangle  = \left[ {1 + \frac{1}{{i\hbar }}\int_{ - \infty }^{ + \infty } {d\tau \;} e^{\frac{i}{\hbar }\hat H_0 \tau } \hat H_{int} e^{ - \frac{i}{\hbar }\hat H_0 \tau } } \right]\left| { \Psi \left( { - \infty } \right)} \right\rangle$ so that the probability amplitude of finding the electron-field system in the eigenstate $\left| {{\bf{P}}_{\rm t} ,E;\xi _1  \ldots \xi _n } \right\rangle$ of energy ${\mathcal E} = E + \sum\nolimits_{j = 0}^n {\hbar \omega _j }$ is $\left\langle {{{\bf{P}}_{\rm t} ,E;\xi _1  \ldots \xi _n }} \mathrel{\left | {\vphantom {{{\bf{P}}_{\rm t} ,E;\xi _1  \ldots \xi _n } {\Psi \left( { + \infty } \right)}}} \right. \kern-\nulldelimiterspace} {{\Psi \left( { + \infty } \right)}} \right\rangle  = A^{\left( 0 \right)}  + A^{\left( 1 \right)}$ where
\begin{eqnarray} \label{amplit}
 A^{\left( 0 \right)}  &=& \delta _{n,0} \bar \psi _0 \left( {{\bf{P}}_{\rm t} ,E} \right), \nonumber \\ 
 A^{\left( 1 \right)}  &=& \frac{{2\pi }}{i}\int {d^2 {\bf{P}}'_{\rm t} } \left\langle {{\bf{P}}_{\rm t} ,E;\xi _1  \ldots \xi _n } \right|\hat H_{int} \left| {{\bf{P}}'_{\rm t} ,{\mathcal E};0} \right\rangle \bar \psi _0 \left( {{\bf{P}}'_{\rm t} ,{\mathcal E}} \right). \nonumber \\ 
\end{eqnarray}
The matrix element inside the integral can be readily evaluated by using the second of Eqs.(\ref{HeHint}) and Eq.(\ref{PtE}), thus getting
\begin{eqnarray}
&& \left\langle {{\bf{P}}_{\rm t} ,E;\xi _1  \ldots \xi _n } \right|\hat H_{int} \left| {{\bf{P}}'_{\rm t} ,{\mathcal E};0} \right\rangle  = \frac{{\delta _{n,1} }}{V}\int {\frac{{d^3 {\bf{R}}}}{{\left( {2\pi \hbar } \right)^3 }} \cdot } \nonumber  \\ 
&\cdot& e^{ - \frac{i}{\hbar }\left( {{\bf{P}}_{\rm t}  - {\bf{P}}'_{\rm t} } \right) \cdot {\bf{R}}_{\rm t} } e^{i\left( {\frac{{\omega _1 }}{V} + \frac{{P_{\rm t}^2  - {P'_{\rm t}}^2 }}{{2\hbar P_0 }}} \right)R_{\rm v} } C^* \left( {{\bf{R}},\xi _1 } \right), 
\end{eqnarray}
which inserted into the second of Eqs.(\ref{amplit}), after using Eqs.(\ref{spect}) and (\ref{envelope}), yields
\begin{eqnarray} \label{A1}
 A^{\left( 1 \right)}  &=& \delta _{n,1} \frac{{\chi ^{\left( {\Delta E} \right)} \left( {E + \hbar \omega _1  - E_0 } \right)}}{{2\pi i\hbar ^2 V}} \int {d^3 {\bf{R}}} e^{ - \frac{i}{\hbar }{\bf{P}}_{\rm t}  \cdot {\bf{R}}_{\rm t}}  \cdot \nonumber  \\ 
  &\cdot& e^{i\left( {\frac{{\omega _1 }}{V} + \frac{{{P_{\rm t}}^2 }}{{2\hbar P_0 }}} \right)R_{\rm v} } C^* \left( {{\bf{R}},\xi _1 } \right)\phi \left( {\bf{R}} \right). 
\end{eqnarray}
Due to the Kronecker deltas, $A^{(0)}$ describes an elastic process where no field excitation is created upon scattering whereas $A^{(1)}$ is related to the inelastic scattering accompanied by the creation of a single field excitation at $\xi_1$. Accordingly the transverse momentum-energy probability distribution of the scattered electron is obtained by summing the contributions of all the field excitations, i.e. ${d \mathcal P}/{({d^2 {\bf{P}}_{\rm t} dE})} = \int {d\xi } | {A^{\left( 1 \right)} } |^2$ (with the relabelling $\xi_1 \rightarrow \xi$) which, after using Eqs.(\ref{A1}) and the second of Eq.(\ref{ChiDelta}) to take the limit $\Delta E \to 0^ +$, yields
\begin{eqnarray} \label{APPdPdPtdE1}
&& \frac{{d \mathcal P}}{{d^2 {\bf{P}}_{\rm t} dE}} = \frac{1}{{\hbar \left( {2\pi \hbar ^2 V} \right)^2 }}\int {d^3 {\bf{r}}} \sum\limits_{j = 1}^3 \cdot \nonumber \\ 
&& \cdot \left| {\int {d^3 {\bf{R}}} e^{ - \frac{i}{\hbar }{\bf{P}}_{\rm t}  \cdot {\bf{R}}_{\rm t}} e^{i\left( {\frac{\omega }{V} + \frac{{P_{\rm t}^2 }}{{2\hbar P_0 }}} \right)R_{\rm v} } \phi \left( {\bf{R}} \right)C^* \left( {{\bf{R}},\xi } \right)} \right|^2
\end{eqnarray}
where the positive frequency $\omega  = \left( {E_0  - E} \right)/\hbar$ has been selected by the delta function. To proceed, we now use the Onsager reciprocity of the Green tensor ${\mathcal G}_\omega  \left( {{\bf{r}},{\bf{r}}'} \right) = {\mathcal G}_\omega ^{\rm T} \left( {{\bf{r}}',{\bf{r}}} \right)$ and the relevant integral identity \cite{Schee}
\begin{equation}
\int {d^3 {\bf{s}}} \left[{\rm Im} \; \varepsilon _\omega \left( {\bf{s}} \right) \right]  {\mathcal G}_\omega  \left( {{\bf{r}},{\bf{s}}} \right){\mathcal G}_\omega ^* \left( {{\bf{s}},{\bf{r}}'} \right) = \frac{{c^2 }}{{\omega ^2 }}{\mathop{\rm Im}\nolimits} \left[ {{\mathcal G}_\omega  \left( {{\bf{r}},{\bf{r}}'} \right)} \right]
\end{equation}
so that, using the expression of $C$ defined after Eq.(\ref{HeHint}) and performing the change of variable $E \rightarrow \omega$, Eq.(\ref{APPdPdPtdE1}) provides the momentum resolved energy loss probability
\begin{eqnarray} \label{APPMoEnLoPr}
 \frac{{d \mathcal P}}{{d^2 {\bf{P}}_{\rm t} d\omega}} &=& \frac{{4\alpha }}{{c }}\int {d^3 {\bf{R}}_1 } \int {d^3 {\bf{R}}_2 } \frac{{e^{-\frac{i}{\hbar }{\bf{P}}_{\rm t}  \cdot \left( {{\bf{R}}_{\rm 1t}  - {\bf{R}}_{\rm 2t} } \right)} }}{{\left( {2\pi \hbar } \right)^2 }} \cdot \nonumber  \\ 
&\cdot& e^{  i\left( {\frac{\omega }{V} + \frac{{P_{\rm t}^2 }}{{2\hbar P_0 }}} \right)\left( {R_{\rm 1v}  - R_{\rm 2v} } \right)} \phi  \left( {{\bf{R}}_1 } \right)\phi ^* \left( {{\bf{R}}_2 } \right) \cdot \nonumber \\
&\cdot& {\mathop{\rm Im}\nolimits} \left[ {{\bf{u}}_{\rm v}  \cdot {\mathcal G}_\omega  \left( {{\bf{R}}_1 ,{\bf{R}}_2 } \right){\bf{u}}_{\rm v} } \right].
\end{eqnarray}
Note that reality of the right hand side of Eq.(\ref{APPMoEnLoPr}) is ensured by the Onsager reciprocity of the Green tensor.

\section{Momentum resolved energy loss probability for a slab}
With reference to Fig.1(a) depicting the inelastic scattering of an aloof electron traveling parallel to a slab, the transverse and longitudinal parts of a vector $\bf F$ are here ${\bf{F}}_{\rm t}  = F_y {\bf{e}}_y  + F_z {\bf{e}}_z$ and ${\bf{F}}_{\rm v}  = F_x {\bf{e}}_x$, respectively. To evaluate the momentum resolved energy loss probability it is convenient to cast Eq.(\ref{MoEnLoPr}) in momentum space by using the envelope representation of Eq.(\ref{envelope}) thus getting
\begin{eqnarray} \label{APPMoReEnLoPr1}
 \frac{{d {\mathcal P}}}{{d^2 {\bf{P}}_{\rm t} d\omega }} &=& \frac{{4\alpha }}{c}\int {d^2 {\bf{P}}_{1 {\rm t}} } \int d^2 {\bf{P}}_{2 {\rm t}} \bar \phi \left( {{\bf{P}}_{1 {\rm t}} } \right)\bar \phi ^* \left( {{\bf{P}}_{2 {\rm t}} } \right) \cdot \nonumber  \\ 
  &\cdot& \int {d^3 {\bf{R}}_1 } \int {d^3 {\bf{R}}_2 } \frac{{e^{ - i{\bf{K}}_1  \cdot {\bf{R}}_1  + i{\bf{K}}_2  \cdot {\bf{R}}_2 } }}{{\left( {2\pi \hbar } \right)^4 }} \cdot  \nonumber \\ 
&\cdot& {\mathop{\rm Im}\nolimits} \left[ {{\bf{e}}_x  \cdot {\cal G}_\omega  \left( {{\bf{R}}_1 ,{\bf{R}}_2 } \right){\bf{e}}_x } \right]
\end{eqnarray}
where the integration over ${\bf R}_1$ and ${\bf R}_2$ provides the double Fourier transform of the imaginary part of the Green tensor component along the electron velocity, evaluated at the two wavevectors ($j=1,2$)
\begin{equation} \label{Kj}
{\bf{K}}_j  = \frac{{{\bf{P}}_{\rm t}  - {\bf{P}}_{j {\rm t}} }}{\hbar } - \left( {\frac{\omega }{V} + \frac{{P_{\rm t}^2  - P_{j {\rm t}}^2 }}{{2\hbar P_0 }}} \right){\bf{e}}_x 
\end{equation}
accounting for the momentum exchange with the sample. In the considered geometry, the slab Green tensor decomposes as 
\begin{eqnarray}
 {\cal G}_\omega  \left( {{\bf{R}}_1 ,{\bf{R}}_2 } \right) &=& {\cal G}_\omega ^{\left( 0 \right)} \left( {{\bf{R}}_1 ,{\bf{R}}_2 } \right) + \Pi \left( {X_1 } \right)\Pi \left( {X_2 } \right) \cdot \nonumber \\ 
  &\cdot& \theta \left( {Z_1 } \right)\theta \left( {Z_2 } \right){\cal G}_\omega ^{\left( r \right)} \left( {{\bf{R}}_1 ,{\bf{R}}_2 } \right),
\end{eqnarray}
where ${\cal G}_\omega ^{\left( 0 \right)} $ and ${\cal G}_\omega ^{\left( r \right)} $ are the free space and reflected parts of the Green tensor of an infinite slab whereas the longitudinal truncation function $\Pi (X)$ has been introduced to model the finite longitudinal length of the slab and it can be any function centered at $X=0$ and satisfying $\int\limits_{ - \infty }^{ + \infty } {dX\;} \Pi ( X ) = L$. The $\theta$-functions have been introduced in the reflected part since the envelopes $\phi \left( {{\bf{R}}_1 } \right)$ and $\phi ^* \left( {{\bf{R}}_2 } \right)$ in Eq.(\ref{MoEnLoPr}) of the aloof electron do not vanish only in vacuum. The free space and reflected parts of the Green tensor of an infinite slab are such that \cite{Cheww}
\begin{eqnarray} \label{G0Gr}
&& {\mathop{\rm Im}\nolimits} [ {\bf{e}}_x  \cdot {\cal G}_\omega ^{\left( 0 \right)} \left( {{\bf{R}}_1 ,{\bf{R}}_2 } \right){\bf{e}}_x  ] =    \nonumber \\
&& =  \int d^3 {\bf{k}}\frac{{e^{i{\bf{k}} \cdot \left( {{\bf{R}}_1  - {\bf{R}}_2 } \right)} }}{{{16\pi ^2 } k_\omega  }} \left( {1 - \frac{{k_x^2 }}{{k_\omega ^2 }}} \right) \delta \left( {k - k_\omega  } \right),  \nonumber \\ 
&& {\mathop{\rm Im}\nolimits} [ {\bf{e}}_x  \cdot  {\cal G}_\omega ^{\left( r \right)} \left( {{\bf{R}}_1 ,{\bf{R}}_2 } \right){\bf{e}}_x  ] =   \nonumber \\
&& =  \int {d^2 {\bf{k}}_\parallel  } \frac{{e^{i{\bf{k}}_\parallel   \cdot \left( {{\bf{R}}_{1\parallel }  - {\bf{R}}_{2\parallel } } \right)} }}{{8\pi ^2 }} {\mathop{\rm Re}\nolimits} \left[ e^{ik_{z0} \left( {Z_1  + Z_2 } \right)}\Upsilon  \right],
\end{eqnarray}
where $k_\omega   = \omega /c$ is the vacuum wavenumber, ${\bf{F}}_\parallel   = F_x {\bf{e}}_x  + F_y {\bf{e}}_y$ is the part of the vector $\bf F$ parallel to the slab, 
\begin{equation} \label{Upsil}
\Upsilon  = \frac{1}{k_{z0}} \left[ \left( {\frac{{ - k_{z\varepsilon }  + k_{z0} }}{{k_{z\varepsilon }  + k_{z0} }}} \right)\frac{{k_y^2 }}{{k_\parallel ^2 }} + \left( {\frac{{k_{z\varepsilon }  - k_{z0} \varepsilon _\omega  }}{{k_{z\varepsilon }  + k_{z0} \varepsilon _\omega  }}} \right)\frac{{k_{z0}^2 }}{{k_\omega ^2 }}\frac{{k_x^2 }}{{k_\parallel ^2 }} \right]
\end{equation}
is the dimensionless factor accounting for the reflection of TE and TM modes and  $k_{z0}  = \sqrt {k_\omega ^2 \left( {1 + i\eta } \right) - k_\parallel ^2 }$ and $k_{z\varepsilon } = \sqrt {k_\omega ^2 \varepsilon_\omega  - k_\parallel ^2 }$ are the normal components of the field wavevectors in vacuum ($0$) and in the slab ($\epsilon_\omega$). Here we have chosen the Riemann sheet of the square roots such that their imaginary part is positive and we have inserted an infinitesimal vacuum absorption ($\eta \rightarrow 0^+$) in $k_{z0}$ for regularization purposes. 

The first of Eqs.(\ref{G0Gr}) states that  ${\rm Im}[{\bf{e}}_x  \cdot {\cal G}_\omega ^{\left( 0 \right)} {\bf{e}}_x]$ only involves the wavevectors of length equal to $k_\omega$ so that its double Fourier transform inside  Eq.(\ref{APPMoReEnLoPr1}) is not vanishing only for those
electron  transverse momenta ${{\bf{P}}_{j {\rm t}} }$ such that $K_j^2  = k_\omega ^2$ which, using Eq.(\ref{Kj}), amounts to  
\begin{equation}
\left( {\frac{{\hbar \omega }}{V} + \frac{{P_{\rm t}^2  - P_{j {\rm t}}^2 }}{{2P_0 }}} \right)^2  = \left( {\frac{{\hbar \omega }}{c}} \right)^2  - \left| {{\bf{P}}_{\rm t}  - {\bf{P}}_{j {\rm t}} } \right|^2. 
\end{equation}
A simple inspection of this equation shows that it can be satisfied by some ${{\bf{P}}_{j {\rm t}} }$ close to ${{\bf{P}}_{\rm t} }$ only if $V$ is very close to $c$ as a consequence of the paraxial requirement $\left| {{\bf{P}}_{j {\rm t}} } \right| \ll P_0$. Therefore, since we are here focused on the sub-relativistic regime, we conclude that free space part of the Green tensor provides no appreciable contribution to Eq.(\ref{APPMoReEnLoPr1}). Inserting the second of Eqs.(\ref{G0Gr}) into Eq.(\ref{APPMoReEnLoPr1}), after some tediuos but straightforward algebra we get
\begin{eqnarray} \label{APPMoReEnLoPr2}
&&  \frac{{d {\mathcal P}}}{{d^2 {\bf{P}}_{\rm t} d\omega }} = \frac{\alpha }{{2\pi ^2 c\hbar ^4 }}{\mathop{\rm Re}\nolimits} \int {d^2 {\bf{k}}_\parallel  } \Upsilon  \cdot \nonumber \\ 
&& \cdot \left[ {\int {d^2 {\bf{P}}_{1 {\rm t}} } \bar \phi \left( {{\bf{P}}_{1 {\rm t}} } \right)\frac{{\tilde \Pi \left( {k_x  - K_{1x} } \right)\delta \left( {k_y  - K_{1y} } \right)}}{{\left( {K_{1z}  - k_{z0} } \right)}}} \right] \cdot \nonumber  \\ 
&& \cdot \left[ {\int {d^2 {\bf{P}}_{2 {\rm t}} \bar \phi ^* \left( {{\bf{P}}_{2 {\rm t}} } \right)} \frac{{\tilde \Pi \left( {k_x  - K_{2x} } \right)\delta \left( {k_y  - K_{2y} } \right)}}{{\left( {K_{2z}  + k_{z0} } \right)}}} \right]
\end{eqnarray}
where $\tilde \Pi \left( K \right) = \frac{1}{{2\pi }}\int\limits_{ - \infty }^{ + \infty } {dX} e^{ - iKX} \Pi \left( X \right)$ is the Fourier transform of the function $\Pi(X)$. Note that Eq.(\ref{APPMoReEnLoPr2}) expresses the momentum resolved energy loss probability as the superposition of the contributions of all the slab optical excitations labelled by their parallel wavevector ${\bf k}_\parallel$. 

In order to have an analytical treatment of such contributions, we choose the envelope waist profile  $\phi \left( {{\bf{R}}_{\rm t} } \right) = \left[ {2/\left( {\pi \sigma ^4 } \right)} \right]^{1/4} \exp \left( { - Y^2 /\sigma ^2 } \right)\exp \left( { - \left| {Z - b} \right|/\sigma } \right)$,  which is a normalized ($\int {d^2 {\bf{R}}_{\rm t} \left| {\phi \left( {{\bf{R}}_{\rm t} } \right)} \right|^2 }  = 1$) Gaussian-exponential profile of width $\sigma \gg \hbar /P_0 $ and impact parameter $b>0$, whose momentum space representation (see Eq.(\ref{phit})) is
\begin{equation} \label{PhitGE}
\bar \phi \left( {{\bf{P}}_{\rm t} } \right) = \left( {\frac{{2\sigma ^4 }}{{\pi ^3 \hbar ^4 }}} \right)^{1/4} \frac{{e^{ - \frac{{\sigma ^2 }}{{4\hbar ^2 }}P_y^2 } e^{ - \frac{i}{\hbar }b P_z } }}{{1 + \frac{{\sigma ^2 }}{{\hbar ^2 }}P_z^2 }}.
\end{equation}
In addition, we choose the slab longitudinal truncation function $\Pi \left( X \right) = e^{ - \frac{2}{L}\left| X \right|}$ whose Fourier transform is
\begin{equation} \label{Pit}
\tilde \Pi \left( K \right) = \frac{L}{{2\pi }}\frac{1}{{1 + \left( {\frac{{KL}}{2}} \right)^2 }}.
\end{equation}
Inserting Eqs.(\ref{PhitGE}) and (\ref{Pit}) into Eq.(\ref{APPMoReEnLoPr2}) we get
\begin{equation}
\frac{{d{\cal P}}}{{d^2 {\bf{P}}_{\rm t} d\omega }} = \frac{{4\alpha \sigma ^2 }}{{c\hbar ^2 \sqrt {2\pi ^3 } }}{\mathop{\rm Re}\nolimits} \left[ \int {d^2 {\bf{k}}_\parallel  } \Upsilon  e^{ - \frac{{\sigma ^2 }}{{2\hbar ^2 }}\left( {P_y  - \hbar k_y } \right)^2 } J^ +  J^ -  \right]
\end{equation}
where ($\tau = \pm 1$)
\begin{eqnarray} \label{APPJtau}
 J^\tau   &=& \left( {\frac{{8\hbar ^4 P_0^2 }}{{\pi L\sigma ^2 }}} \right)\frac{1}{{2\pi i}}\int\limits_{ - \infty }^{ + \infty } {dP} \;\frac{{e^{\frac{i}{\hbar }Pb} }}{{\left[ {P - \left( {\hbar k_{z0}  + \tau P_z } \right)} \right]}} \cdot \nonumber \\ 
&\cdot& \frac{1}{{\left( {P^2  + \frac{{\hbar ^2 }}{{\sigma ^2 }}} \right)\left( {P^2  - q_ + ^2 } \right)\left( {P^2  - q_ - ^2 } \right)}} 
 \end{eqnarray}
and 
\begin{equation}
q_ \pm   = \sqrt {2\hbar P_0 \left[ {\left( {k_x  + \frac{\omega }{V}} \right) \pm \frac{{2i}}{L}} \right] + P_{\rm t}^2  - \left( {P_y  - \hbar k_y } \right)^2 }.
\end{equation}
where we have chosen the Riemann sheet of the square root such that its imaginary part is positive. Now the integral in Eq.(\ref{APPJtau}) can be readily evaluated by resorting to the Jordan's lemma so that, after noting that only the residues of poles in the upper half P-plane contribute to the integral since $b>0$, we get  
\begin{eqnarray}
J^\tau   = F_0^\tau  \;e^{i\left( {\tau \frac{{P_z }}{\hbar } + \sqrt {\frac{{\omega ^2 }}{{c^2 }} - k_\parallel ^2 } } \right)b}  + F_\sigma ^\tau  \;e^{ - \frac{b}{\sigma }}  + F_ + ^\tau  e^{\frac{i}{\hbar }q_ +  b}  + F_ - ^\tau  e^{\frac{i}{\hbar }q_ -  b} \nonumber \\ 
\end{eqnarray}
where 
\begin{eqnarray} \label{F}
 F_0^\tau   &=& \frac{{\tilde \Pi \left( {k_x  + \frac{\omega }{V} + \frac{{P_{\rm t}^2  - \left( {P_y  - \hbar k_y } \right)^2  - \left( {\tau P_z  + \hbar k_{z0} } \right)^2 }}{{2\hbar P_0 }}} \right)}}{{\left[ {1 + \sigma ^2 \left( {k_{z0}  + \tau \frac{{P_z }}{\hbar }} \right)^2 } \right]}}, \nonumber \\ 
 F_\sigma ^\tau   &=&   - \frac{{\tilde \Pi \left( {k_x  + \frac{\omega }{V} + \frac{{P_{\rm t}^2  - \left( {P_y  - \hbar k_y } \right)^2  + \frac{{\hbar ^2 }}{{\sigma ^2 }}}}{{2\hbar P_0 }}} \right)}}{{2\left[ {1 + i\sigma \left( {k_{z0}  + \tau \frac{{P_z }}{\hbar }} \right)} \right]}}, \nonumber \\ 
F_ \pm ^\tau   &=&  \pm \frac{i}{{2\pi }}\frac{{P_0 }}{{\left[ {\left( {k_{z0}  + \tau \frac{{P_z }}{\hbar }} \right) - \frac{{q_ \pm  }}{\hbar }} \right]\left( {1 + \frac{{\sigma ^2 }}{{\hbar ^2 }}q_ \pm ^2 } \right)q_ \pm  }}. \nonumber \\
\end{eqnarray}

\end{document}